\documentclass[instruments,article,accept,pdftex,oneauthor]{Definitions/mdpi} 

\firstpage{1} 
\pubvolume{1}
\issuenum{1}
\articlenumber{0}
\pubyear{2024}
\copyrightyear{2024}
\externaleditor{Academic Editor: Nicolas Delerue}
\datereceived{14 December 2023} 
\daterevised{19 January 2024} 
\dateaccepted{21 February 2024} 
\datepublished{ } 
\hreflink{https://doi.org/} 

\Title{A Modified Slicing Method with Multi-Dimensional Unfolding to Measure Hadron-Argon Cross Sections}

\TitleCitation{A Modified Slicing Method with Multi-Dimensional Unfolding to Measure Hadron-Argon Cross Sections}

\Author{{Yinrui Liu} 
 \orcidA{}}

\AuthorNames{Yinrui Liu}

\AuthorCitation{{Liu, Y.} 
}

\address[1]{%
 {Department of Physics, }University 
 of Chicago, Chicago, IL 60637, USA; yinrui@uchicago.edu}

\abstract{Liquid argon technology is widely used by many previous and current neutrino experiments, and it is also promising for future large-scale neutrino experiments. When detecting neutrinos using liquid argon, many hadrons are involved, which can also interact with argon nuclei. In order to gain a better understanding of the detection processes, and to simulate neutrino events, knowledge of hadron-argon cross sections is needed. This paper describes a new procedure which improves upon the previous work with multi-dimensional unfolding to measure hadron-argon cross sections in a liquid argon time projection chamber. Through a simplified version of simulation, we demonstrate the validity of this procedure.}


\begin{document}


\section{Introduction}
Liquid argon (LAr) technology is widely used by many previous and current neutrino experiments, such as MicroBooNE~\cite{MicroBooNE:2016pwy}, ArgoNeuT~\cite{Anderson:2012vc}, and ICARUS~\cite{ICARUS:2004wqc}, and it is also planned to be employed by future experiments such as SBND~\cite{Machado:2019oxb} as well as one of the next-generation large-scale neutrino experiments DUNE~\cite{DUNE:2016hlj}. Neutrinos are mostly detected by their interactions with argon nuclei, in which many types of hadrons are involved, including both the knockout of nucleons and the production of mesons. These neutrino-induced hadrons can also interact with nucleons before they escape the nucleus. This may change the kinematics and types of final state particles that are detected, which complicates the reconstruction of neutrino interactions. These are known as final state interaction (FSI) effects. In addition, the propagation and further interactions of these final state hadrons also need to be simulated properly. Therefore, knowledge of hadron-argon cross sections is required, which is useful for informing FSI and improve simulations as well as their associated uncertainties. 

However, there are few experimental data available on argon, and the predictions are mostly derived by interpolating cross section results from lighter and heavier \mbox{nuclei~\cite{Dytman:2021ohr,PinzonGuerra:2018rju},} such as carbon, sulfur, and iron, which have more data available~\cite{Wilkin:1973xd,Clough:1974qt,Carroll:1976hj,Ashery:1981tq}. In those experiments, the common set-up was to implement a beam of a certain type of hadron of interest and shoot the beam toward a thin target of the material of interest. The survival rate of the hadron beam after the thin target can be measured and used to calculate the cross section. The increasing popularity of LAr-based detectors has motivated efforts toward making cross section measurements of LAr. The LArIAT collaboration proposed the thin slice method to measure hadron-argon cross sections using an LAr time projection chamber (LArTPC)~\cite{LArIAT:2021yix}, which itself can no longer be considered a thin target of LAr for hadrons. The precise track reconstruction {capability of LArTPCs} enables researchers to hypothetically divide the detector into several thin slices, and each slice can be considered an {individual thin target experiment. The measurements from all these slices fill the corresponding energy bins.}

The original method treats the measured cross section in each bin independently and performs an effective correction in each bin to account for inefficiency and bin migration caused by the detector's resolution. We keep the essential idea of the thin slice method and further develop the method with more rigorous statistical procedures, including using multi-dimensional unfolding to consider the full correlations of different measurements. In this paper, Section~\ref{sec:Cross-section formula} shows the derivation of the cross section formula, and Section~\ref{sec:Slicing method} describes the slicing method {in more detail}. Section~\ref{sec:Procedures and results} describes the measurement procedures for a simplified simulation sample, where all results are derived using an {IPython notebook,} 
 referred to as {\texttt{hadron-Ar\_XS}} 
 \cite{hadron-Ar_XS}. Some further discussions on the results as well as {a} summary are given in Section~\ref{sec:Discussions and summary}.

\section{Cross Section Formula}
\label{sec:Cross-section formula}
The total cross section $\sigma_{\rm total}$ as a function of the incident particle's kinetic energy $E$\footnote{For convenience, $\sigma$ and $\sigma(E)$ are used interchangeably throughout the text.} is defined according to
\begin{equation}
    \frac{d\Phi}{dx}=-n\sigma_{\rm total}\Phi,
\label{eqn:dPhidx}
\end{equation}
where $\Phi$ denotes the particle beam flux, $d\Phi$ is the infinitesimal reduction of flux, and $n$ is the number density of the target material. By moving $dx$, the infinitesimal path length of the particle inside the material shown on the right-hand side of Equation~\eqref{eqn:dPhidx}, and then integrating both sides, \mbox{we obtain}
\begin{equation}
    \Phi=\Phi_0e^{-n\sigma_{\rm total}\delta x},
\end{equation}
where $\delta x$ is the path length integral. This assumes that the cross section $\sigma$ remains constant within the variation of $E$ during its passage of $\delta x$.\footnote{In reality, $\sigma$ indicates an effective mean value for cross section within the variation of $E$, since there will always be energy loss during the particle's passage inside the material when we measure the cross section. This also applies to a finite passage length, as discussed in the last paragraph of the section.} For a certain area and a certain period of time, the number of surviving particles detected is proportional to the outgoing particle flux, and thus we have
\begin{equation}
    e^{-n\sigma_{\rm total}\delta x}=\frac{\Phi}{\Phi_0}=\frac{N_{\rm surviving}}{N_{\rm incident}}.
\end{equation}
We can also define the number of interacting particles as $N_{\rm interacting}=N_{\rm incident}-N_{\rm surviving}$. Therefore, after measuring the number of incident particles and the number of surviving particles, the total cross section can be calculated as follows:
\begin{equation}
    \sigma_{\rm total}=-\frac{1}{n\delta x}\ln\left(\frac{N_{\rm surviving}}{N_{\rm incident}}\right)=-\frac{1}{n\delta x}\ln\left(\frac{N_{\rm incident}-N_{\rm interacting}}{N_{\rm incident}}\right).
\label{eqn:XStot}
\end{equation}

When it comes to the exclusive cross section\footnote{Even for the inclusive cross section, there may be a reduction in flux due to particle decay. In this case, $a$ denotes the total inelastic interaction, $b$ denotes the particle decay, and thus $\sigma_b$ is considered an effective cross section. For convenience, we will also refer to particle decay as an ``interaction''.}, we denote $a$ as the signal interaction and $b$ as all the other interactions, and thus we have $d\Phi=d\Phi_a+d\Phi_b$, where $d\Phi_a$ is the reduction in flux due to the signal interaction. Also, we have $\sigma_{\rm total}=\sigma_a+\sigma_b$. By separating $N_{\rm interacting}$ based on the type of interactions into $N_{\rm interacting}^a+N_{\rm interacting}^b$ in Equation~\eqref{eqn:XStot}, we obtain
\begin{equation}
    \sigma_a+\sigma_b=-\frac{1}{n\delta x}\ln\left(1-\frac{N_{\rm interacting}}{N_{\rm incident}}\right)=-\frac{1}{n\delta x}\ln\left(1-\frac{N_{\rm interacting}^a+N_{\rm interacting}^b}{N_{\rm incident}}\right).
    \label{eqn:XS_a_b}
\end{equation}
where $\sigma_a$ and $\sigma_b$ in Equation~\eqref{eqn:XS_a_b} are not separable given the logarithm function on the right-hand side. Only when $N_{\rm interacting}\ll N_{\rm incident}$, which implies that $\delta x$ is quite small, and the thin target approximation holds {can we} use the approximation $\lim_{x\to0}\ln(1-x)=-x$ and find
\begin{equation}
    \sigma_a+\sigma_b=\frac{1}{n\delta x}\frac{N_{\rm interacting}^a+N_{\rm interacting}^b}{N_{\rm incident}}.
\end{equation}
Therefore, we have 
\begin{equation}
    \sigma_a=\frac{1}{n\delta x}\frac{N_{\rm interacting}^a}{N_{\rm incident}}\,({\rm and}\,\sigma_b=\frac{1}{n\delta x}\frac{N_{\rm interacting}^b}{N_{\rm incident}}),
    \label{eqn:XS_thin}
\end{equation}
which is in fact a direct implication from the definition of the exclusive cross section
\begin{equation}
    \frac{d\Phi_a}{dx}=-n\sigma_a\Phi.
\label{eqn:dPhiadx}
\end{equation}

However, in the slicing method described in Section~\ref{sec:Slicing method}, $\delta x$ for each slice, which we used to calculate $\sigma$, is not necessarily small. Therefore, we wish to obtain an unbiased cross section formula without the thin target approximation. From Equations~\eqref{eqn:dPhidx} and \eqref{eqn:dPhiadx}, we have 
\begin{equation}
    \frac{\sigma_{\rm total}}{\sigma_a}=\frac{d\Phi}{d\Phi_a}.
\end{equation}
For a finite $\delta x$, we can estimate this relationship as 
\begin{equation}
    \frac{\overline{\sigma}_{\rm total}}{\overline{\sigma}_a}=\frac{\int d\Phi}{\int d\Phi_a}=\frac{\Delta\Phi}{\Delta\Phi_a}=\frac{N_{\rm interacting}}{N_{\rm interacting}^a},
\end{equation}
where $\overline{\sigma}$ is the effective mean value for the cross section within the variation of $E$ during the passage of $\delta x$.\footnote{According to Equation~\eqref{eqn:dPhidx} and \eqref{eqn:dPhiadx}, assuming $n$ as constant, $\bar{\sigma}$ can be expressed as $\int_{\delta x}\Phi\sigma dx$.} Therefore, combined with Equation~\eqref{eqn:XStot}, we have the expression for any channel $a$:
\begin{equation}
    \overline{\sigma}_a=\frac{N_{\rm interacting}^a}{nN_{\rm interacting}\delta x}\ln\left(\frac{N_{\rm incident}}{N_{\rm incident}-N_{\rm interacting}}\right).
\label{eqn:XS_spatial}
\end{equation}
Because we can never measure $\sigma$ in an infinitely small $E$ bin, we will express $\overline{\sigma}$ as $\sigma$ in the following sections. In the thin target approximation, where $N_{\rm interacting}\ll N_{\rm incident}$, Equation~\eqref{eqn:XS_spatial} can be approximated to Equation~\eqref{eqn:XS_thin}.

\section{Slicing Method}
\label{sec:Slicing method}
An LArTPC cannot be seen as a thin target in terms of hadrons, whose mean free path in LAr is to the order of 10--100 cm. However, thanks to its high-resolution track reconstruction ability, the LArIAT collaboration proposed the thin slice method~\cite{LArIAT:2021yix}, where the detector is hypothetically divided into several slices along the hadron beam direction. Each slice is viewed as a thin target with a width of several millimeters, based on the spacing of the sensing wires. When detecting tracks in the TPC, each slice serves as an {individual} thin target experiment. By detecting where the track ends, we know where the interaction happens and thus fill in the corresponding energy bins of $N_{\rm interacting}$ and $N_{\rm incident}$, which are used to calculate the cross section. The final results are rebinned to wider energy bins, such as 50 MeV, in order to obtain the results.

{Based on the thin slice method, Stocker {et al.} first proposed the idea of energy slicing}~\cite{fStocker} based on a study of the ProtoDUNE-SP experiment~\cite{DUNE:2020cqd}. {In the energy slicing method, each energy bin is directly considered a slice, which is natural since the cross section is measured as a function of the kinetic energy of the incident particle.} \footnote{The authors of~\cite{Liu:2023kfh} also had a similar description of the slicing method, while the cross section formula used in these papers is proven to be an approximation of Equation~\eqref{eqn:XS_spatial}, according to Section~\ref{sec:Cross-section formula}.} Figure~\ref{fig:slicing_method} shows an illustration of an LArTPC. A beam hadron is incident from the left side of the detector and leaves a track inside the detector. The beam hadron track ends at the end vertex, where either an interaction occurs or the hadron comes to rest, potentially producing some daughter particles, which can be used to determine the type of interaction{ or the decay}. The kinetic energy of the beam hadron when it enters the detector is denoted by $E_{\rm initial}$, which is known from the beam, and it approximately follows a Gaussian distribution, given the momentum spread. The kinetic energy of the beam hadron at the end vertex is denoted by $E_{\rm end}$. Given these two energies, the track can be divided into several slices based on the predefined energy bins. {\footnote{From this point onward, the slice can be used interchangeably with the energy bin.} The bin edges are indicated by dark red bars in Figure~\ref{fig:slicing_method}, where the last bar is dashed because the beam hadron did not reach that energy. As shown in Figure~\ref{fig:slicing_method}, the first complete slice is referred to as the initial slice, and the slice which has the end vertex is referred to as the end slice. If the interaction occurring at the end vertex is a signal interaction, then the end slice is also referred to as the interaction slice.
The piece of track prior to the initial slice is referred to as an incomplete slice, which will not be used. {In contrast}, $E_{\rm end}$ is inside the end slice. 
\begin{figure}[H]
    \centering
    \includegraphics[width=0.8\columnwidth]{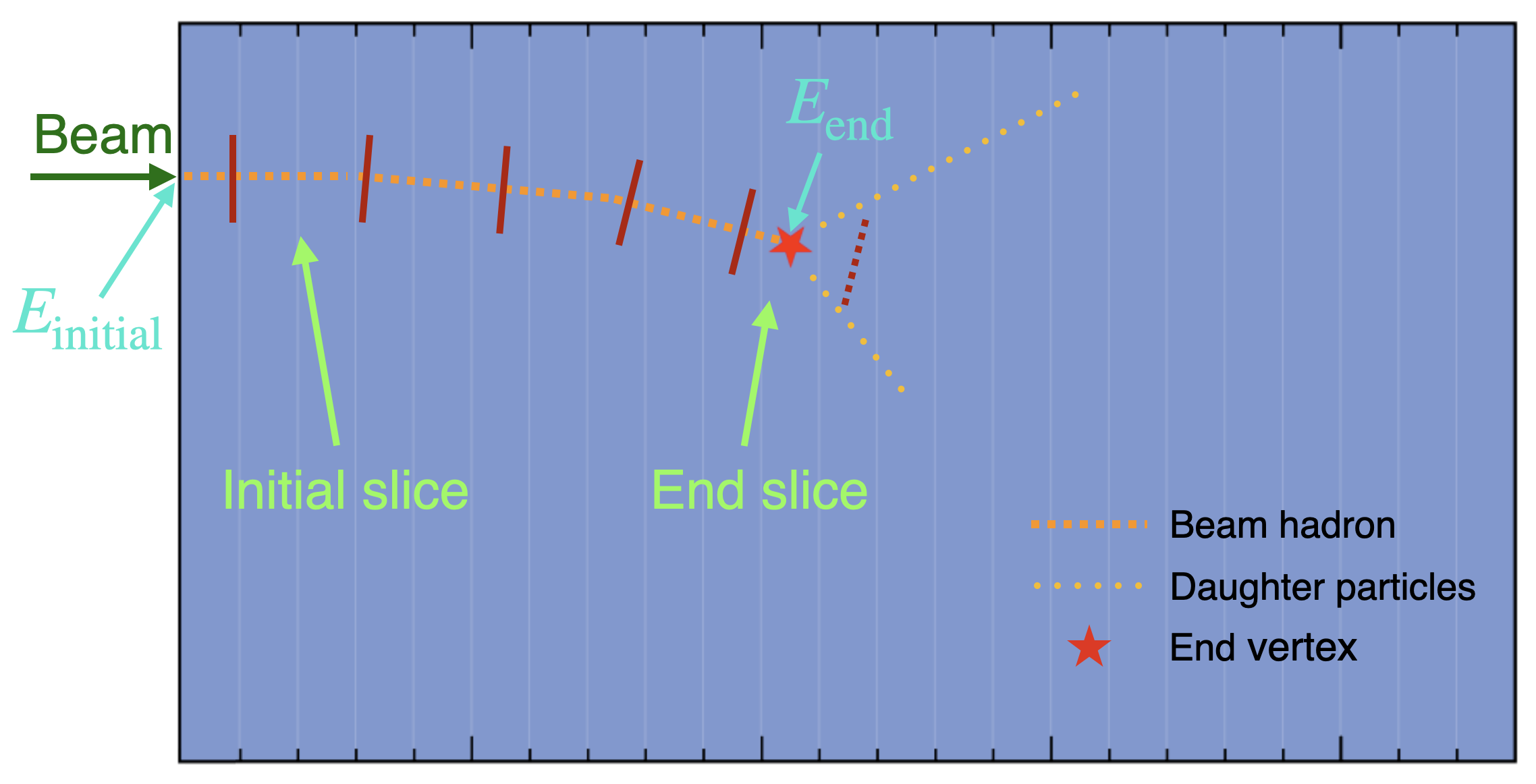}
    \caption{An illustration of an LArTPC, where a beam hadron is shot into the detector from the left side. More descriptions of the elements in the illustration are provided in the text.}
    \label{fig:slicing_method}
\end{figure}

For convenience, we define the slice index ${\rm ID}$ from 1 to the number of energy bins $N$, starting with the highest energy bin. Therefore, for each beam hadron track, there is an initial slice index ${\rm ID_{ini}}$, an end slice index ${\rm ID_{end}}$, as well as an interaction slice index ${\rm ID_{int}}$, which is designated as null if the interaction occurring at the end vertex is not the signal interaction. In addition, if the end vertex is inside the incomplete slice, then the whole track is not usable, and thus the indices for all three slices will be designated as null. For a sample of events with a beam hadron track in the detector, the distribution of ${\rm ID_{ini}}$ forms the initial histogram $N_{\rm ini}({\rm ID})$, and similarly, we have the end histogram $N_{\rm end}({\rm ID})$ and the interaction histogram $N_{\rm int}({\rm ID})$. We also define the incident histogram $N_{\rm inc}({\rm ID})$, which will later appear in the cross section formula in Equation~\eqref{eqn:XS_energy}. Each bin of $N_{\rm inc}({\rm ID})$ counts the number of tracks which reach the energy corresponding to the slice index ${\rm ID}$, and thus we say the tracks are incident to that slice. Note that for $N_{\rm inc}({\rm ID})$, one event is likely to contribute to multiple bins, since a track can be incident to a sequence of slices until it interacts. In the thin slice method, for each track, {${\rm ID_{ini}}$ fills $N_{\rm ini}({\rm ID})$, and ${\rm ID_{end}}$ fills $N_{\rm end}({\rm ID})$, while $N_{\rm inc}({\rm ID})$ should be filled from the value of ${\rm ID_{ini}}$ to ${\rm ID_{end}}$. Equivalently, $N_{\rm inc}({\rm ID})$ can also be calculated by the derived $N_{\rm ini}({\rm ID})$ and $N_{\rm end}({\rm ID})$ as shown below:~\cite{fStocker} \footnote{This calculation enables us to derive $N_{\rm inc}({\rm ID})$ using the unfolded histogram given in Section~\ref{sec:Measurement effects}. This is because after unfolding, the event-wise information is lost, and $N_{\rm inc}({\rm ID})$ cannot be derived by counting events. In addition, for $N_{\rm inc}({\rm ID})$, it is no longer one entry per track, and thus it would be problematic to unfold the counted $N_{\rm inc}({\rm ID})$ directly.}
\begin{equation}
\begin{split}
    N_{\rm inc}({\rm ID})=\sum_{j={\rm ID}}^{N}N_{\rm end}(j)-\sum_{j={\rm ID}+1}^{N}N_{\rm ini}(j),\\
    {\rm or}\,\,N_{\rm inc}({\rm ID})=\sum_{j=1}^{{\rm ID}}N_{\rm ini}(j)-\sum_{j=1}^{{\rm ID}-1}N_{\rm end}(j).
    \label{eqn:Ninc}
\end{split}
\end{equation}
The two expressions are equivalent given that
\begin{equation}
    \sum_{{\rm ID}=1}^NN_{\rm ini}({\rm ID})=\sum_{{\rm ID}=1}^NN_{\rm end}({\rm ID}),
\end{equation}
which equals the total number of beam hadron events. Given the relationship between the slice index ${\rm ID}$ and the energy $E$ by definition, all of these histograms can also be given as energy histograms.

Compared with Equation~\eqref{eqn:XS_spatial}, by replacing $N_{\rm incident}$ with $N_{\rm inc}(E)$, $N_{\rm interacting}$ with $N_{\rm end}(E)$, $N_{\rm interacting}^a$ with $N_{\rm int}(E)$, and also $\frac{1}{\delta x}$ with $\frac{1}{\delta E}\frac{dE}{dx}$, we derive the cross section for the signal interaction in each energy bin, given by 
\begin{equation}
    \sigma(E)=\frac{N_{\rm int}(E)}{nN_{\rm end}(E)\delta E}\frac{dE}{dx}(E)\ln\left(\frac{N_{\rm inc}(E)}{N_{\rm inc}(E)-N_{\rm end}(E)}\right),
    \label{eqn:XS_energy}
\end{equation}
where $\delta E$ is the energy bin width and $\frac{dE}{dx}(E)$ is the stopping power of the hadron in LAr. Therefore, for each beam hadron event, three properties are needed, which are $E_{\rm initial}$, $E_{\rm end}$, and whether or not there is signal interaction in order to derive the slice indices for $N_{\rm ini}$, $N_{\rm end}$, and $N_{\rm int}$. This allows us to treat the three indices as a combined 3D variable, thus enabling the multi-dimensional unfolding discussed in Sections~\ref{sec:Deriving the statistical uncertainty} and \ref{sec:Measurement effects}.

\section{Procedures and Results}
\label{sec:Procedures and results}
{This section describes the detailed procedures for measuring the hadron-argon cross section with the help of the IPython notebook} \texttt{hadron-Ar\_XS}~\cite{hadron-Ar_XS}. We first describe how the simulation samples are prepared (Section~\ref{sec:Simulations}) and then use their true information to extract the true cross sections (Section~\ref{sec:Extracting the true cross section}). The derivation of statistical uncertainty is described in Section~\ref{sec:Deriving the statistical uncertainty}. Section~\ref{sec:Measurement effects} talks about how we model the measurement effects and prepare the fake data sample. Finally, the measured cross section results of the fake data sample are presented in Section~\ref{sec:Fake data results}.
\subsection{Simulations}
\label{sec:Simulations}
All results presented in this paper were obtained from data simulated in \texttt{hadron-Ar\_XS}~\cite{hadron-Ar_XS}. Although this paper focuses on the method for calculating the cross section, it is worth describing how the simulation was carried out. Smooth and positive curves were created to represent the hadron-argon cross sections $\sigma$ as functions of the hadron's kinetic energy $E$, as shown in Figure~\ref{fig:XScurves}. The signal cross section $\sigma_{\rm sig}(E)$ was the cross section for an exclusive channel that we wanted to measure, and other cross sections $\sigma_{\rm oth}(E)$ accounted for all the others. The total cross section is given as $\sigma_{\rm tot}(E)=\sigma_{\rm sig}(E)+\sigma_{\rm oth}(E)$.

As described in Section~\ref{sec:Slicing method}, for each beam hadron event, we needed its initial kinetic energy $E_{\rm ini}$, the kinetic energy at the end vertex $E_{\rm end}$, as well as the type of interaction occurring at the end vertex in order to use the slicing method to measure the cross section. Therefore, in our simplified simulation, we only aimed to generate these three properties for each event. For $E_{\rm ini}$, we generated a random value following a Gaussian distribution for each event in order to mimic the momentum spread. {\footnote{This assumes the momentum has a central value, whereas most test-beam experiments have multiple momentum modes. Linearly combining data of different momentum modes will not affect the effectiveness of the procedures.} For the latter two, we simulated the hadron's passage inside LAr with a customized step size $\Delta x$. \footnote{The simulation used in this paper was generated with $\Delta x=0.1$ cm, which should be much smaller than the mean free path of a particle with a cross section to the order of hundreds of millibarns.} This means that in each step, we generated a random indicator and decided whether the signal interaction or other interactions happened. If not, then it proceeded to the next step until the hadron reacted or its kinetic energy reached zero. Thus, it also included the simulation of particle energy loss inside LAr. We used the Bethe--Bloch formula~\cite{Workman:2022ynf} to model the mean $dE/dx$ curve as a function of the hadron kinetic energy. In each step, a random value was generated following a simplified version of the Landau--Vavilov distribution~\cite{Workman:2022ynf} as the $dE/dx$ value to be used in the step. {The mean $dE/dx$ value of the Landau--Vavilov distribution employed in each step aligned with the value calculated by the Bethe--Bloch formula at the kinetic energy in that step.} As a result, the energy loss in each step would be $\Delta E = dE/dx\cdot \Delta x$ if no interaction occurred. The mean $dE/dx$ curve used in the simulation and an example $dE/dx$ distribution derived at $E=400$ MeV, where the mean $dE/dx$ was approximated to be 2.10 MeV/cm, are shown in Figure~\ref{fig:dEdx}.
\begin{figure}[H]
    \centering
    \includegraphics[width=0.9\columnwidth]{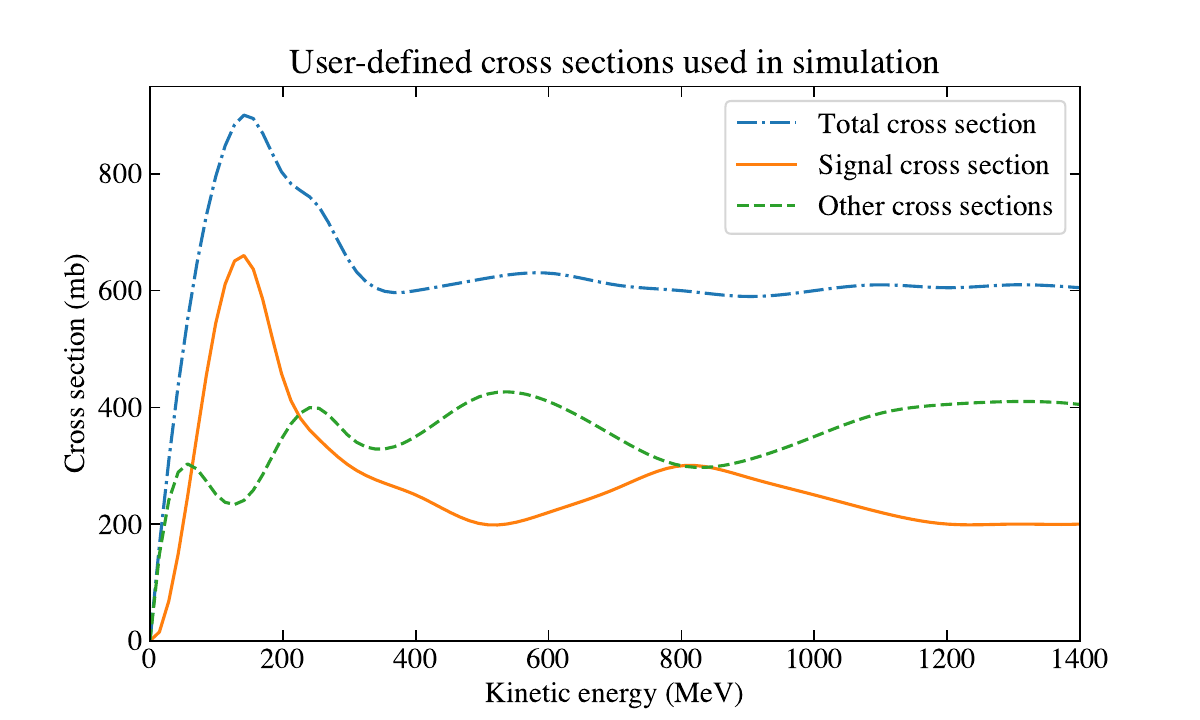}
    \caption{Cross section curves based on which the simulation was generated. The total cross section (blue dash-dotted curve) is the sum of the signal cross section (orange solid curve) and the other cross sections (green dashed curve). {The curves do not correspond to any real hadrons in LAr, but they were created to have an order of magnitude in the hundreds of millibarns, which is similar to the real case. The signal cross section curve also imitates a $\Delta$ resonant peak at about 200 MeV in the low-energy region.}}
    \label{fig:XScurves}
\end{figure}
\begin{figure}[H]
    \centering
    \includegraphics[width=0.48\columnwidth]{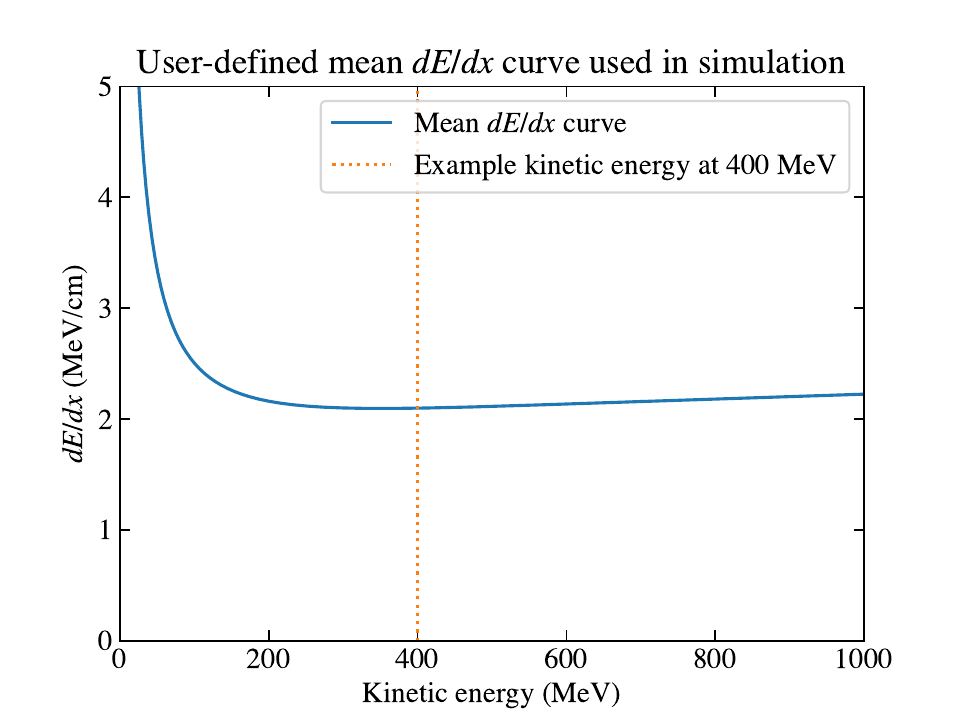}\put(-185,130){(a)}
    \includegraphics[width=0.48\columnwidth]{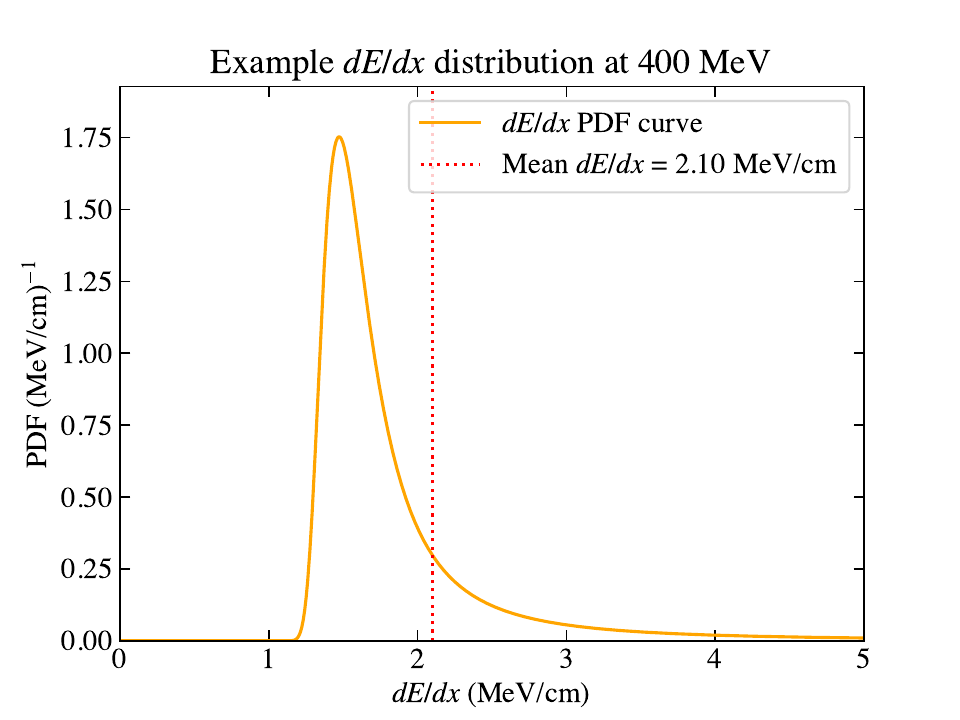}\put(-185,130){(b)}
    \caption{{(\textbf{a}) The} 
 mean $dE/dx$ curve used in the simulation. The dashed vertical line at \mbox{$E=400$ MeV} indicates the case of the example $dE/dx$ distribution. (\textbf{b}) The example $dE/dx$ distribution at \mbox{$E=400$ MeV,} where the mean $dE/dx$ is approximately 2.10 MeV/cm{, which aligns with the value given in subplot (\textbf{a})}.}
    \label{fig:dEdx}
\end{figure}
A simulation sample of 100,000 events was generated. In this simplified simulation, each event had only three properties relevant to the cross section calculations, which were $E_{\rm ini}$, $E_{\rm end}$, and a flag indicating the fate of the hadron. The distributions of these three properties for the simulation sample are shown in Figure~\ref{fig:3properties}. \footnote{There could be a fourth property for each event, which is the event weight. It can be useful when reweighting the simulation sample to study the systematic uncertainties~\cite{Calcutt:2021jaz,Diurba:2021bac}. To simplify the problem, we assigned uniform weights to all samples used in this paper, but the procedures also applied to the samples with non-uniform weights.}
\begin{figure}[H]
    \centering
    \includegraphics[width=0.48\columnwidth]{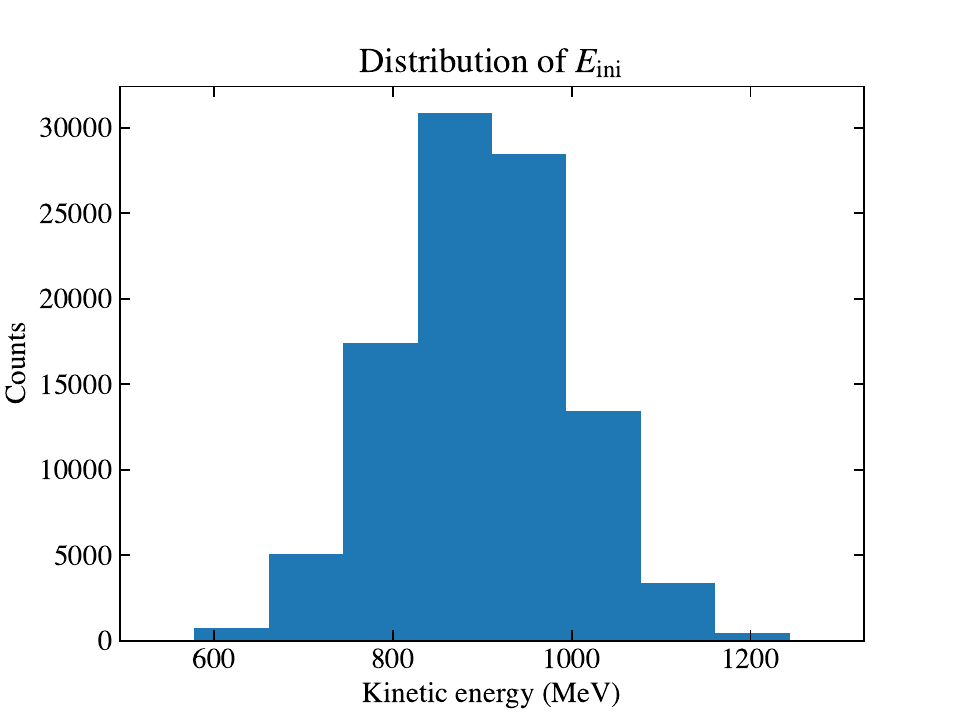}\put(-185,130){(a)}
    \includegraphics[width=0.48\columnwidth]{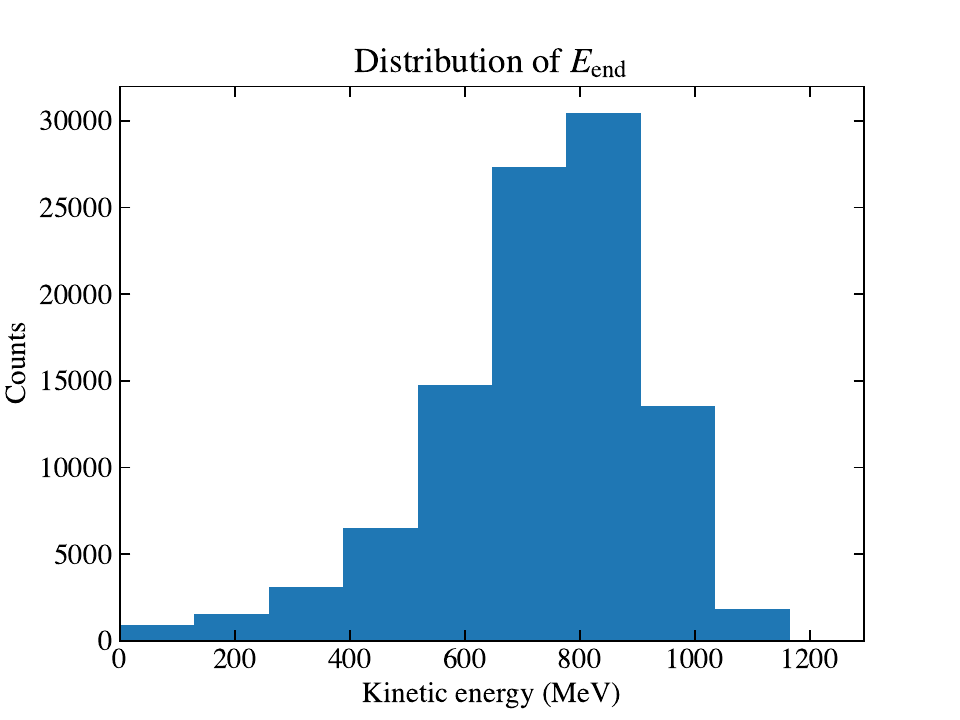}\put(-185,130){(b)}
    
    \centering{\includegraphics[width=0.48\columnwidth]{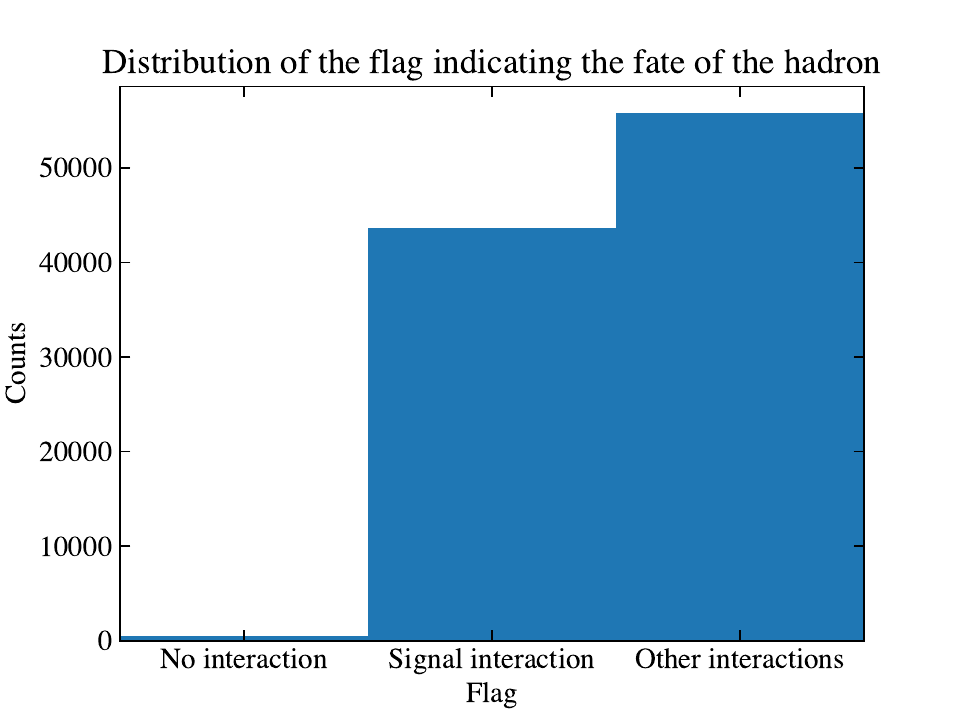}\put(-185,130){(c)}}
    \caption{For the simulation sample, the distribution of (\textbf{a}) $E_{\rm ini}$, (\textbf{b}) $E_{\rm end}$, and (\textbf{c}) the flag indicating the fate of the hadron, where it either has no interaction before it comes to rest, or it has signal interaction or other~interactions.}
    \label{fig:3properties}
\end{figure}

\subsection{Extracting the True Cross Section}
\label{sec:Extracting the true cross section}
From the three properties associated with each event, we used an even binning with $\delta E=50$ MeV\footnote{The binning did not need to be even, and this should be decided on a case-by-case basis.} and obtained the relevant histograms as described in Section~\ref{sec:Slicing method}: $N_{\rm initial}$ as the distribution of $E_{\rm ini}$, $N_{\rm end}$ as the distribution of $E_{\rm end}$, and $N_{\rm interaction_{ex}}$ as the distribution of $E_{\rm end}$, but only for events having the signal interaction. After that, we calculated $N_{\rm incident}$ using Equation~\eqref{eqn:Ninc}. The obtained histograms are shown in Figure~\ref{fig:Nhist}.

By inserting these histograms into Equation~\eqref{eqn:XS_energy}, we derived the signal cross section $\sigma_{\rm sig}(E)$\footnote{In principle, $\sigma_{\rm oth}(E)$ and $\sigma_{\rm tot}(E)$ can also be derived using the same method.}, as shown in Figure~\ref{fig:trueXS}. This cross section was calculated originally using the true values of the three properties of each event, and its consistency with the simulation curve suggests the feasibility of the slicing method. For a quantitative comparison of the extracted cross section from the simulation sample against the input curve, we can calculate $\chi^2$, given as 
\begin{equation}
    \chi^2=\left[\sigma-\sigma_{\rm curve}\right]\cdot V_{\sigma}^{-1}\cdot \left[\sigma-\sigma_{\rm curve}\right]^T,
    \label{eqn:chi2}
\end{equation}
where $\sigma_{\rm curve}$ is a vector of the input cross section evaluated at the middle point in each energy bin and $V_\sigma$ denotes the covariance matrix for the calculated true cross sections $\sigma$. The derivation of $V_\sigma$ is described later in Section~\ref{sec:Deriving the statistical uncertainty}. {The p value shown in the legend in Figure}~\ref{fig:trueXS} suggests good consistency of the sample with the curve. In Section~\ref{sec:Discussions and summary}, toy studies are performed to further validate the slicing method.
\begin{figure}[H]
    \centering
    \includegraphics[width=0.48\columnwidth]{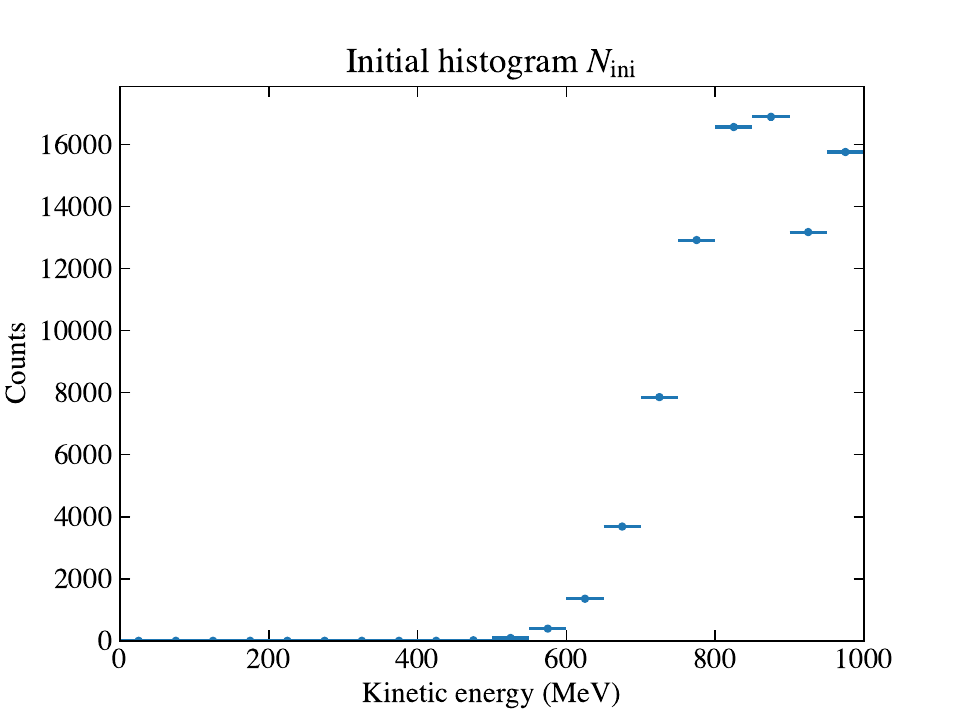}\put(-185,130){(a)}
    \includegraphics[width=0.48\columnwidth]{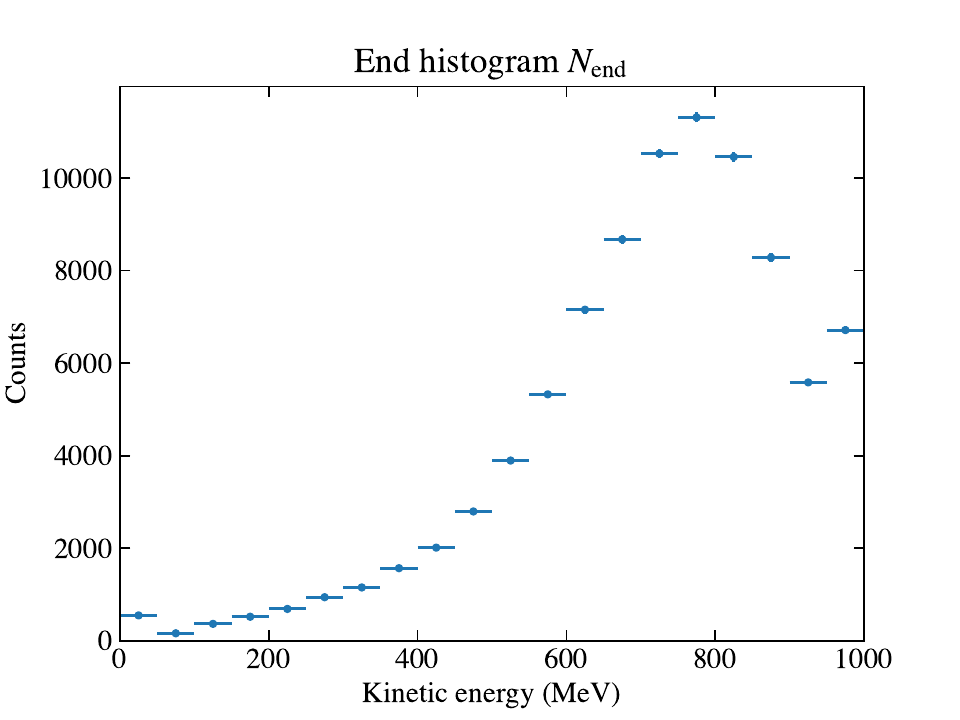}\put(-185,130){(b)}
    
    \includegraphics[width=0.48\columnwidth]{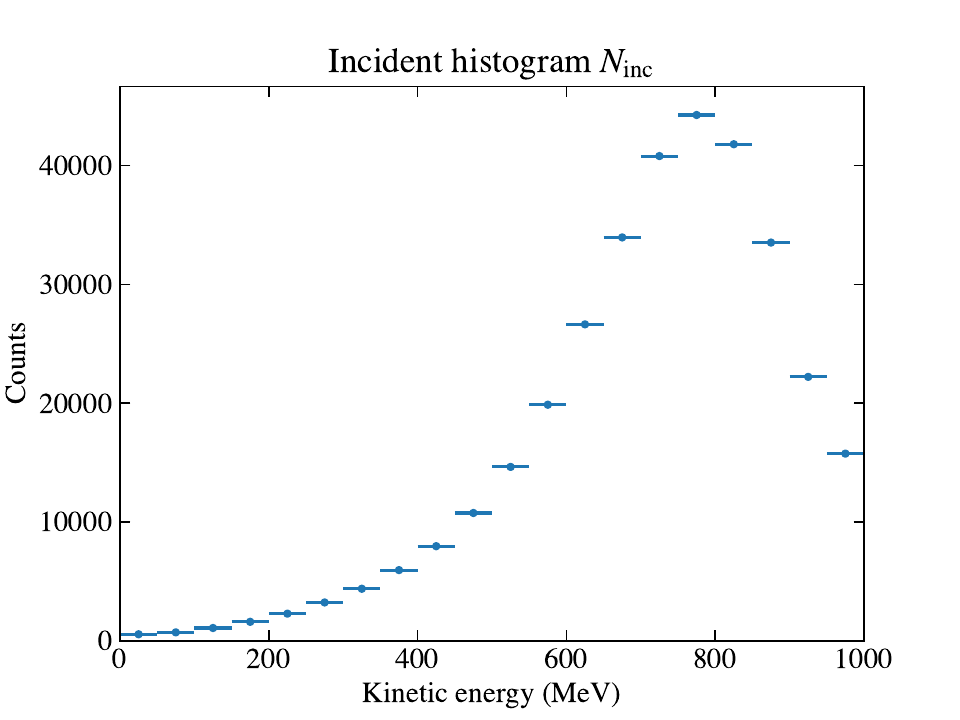}\put(-185,130){(c)}
    \includegraphics[width=0.48\columnwidth]{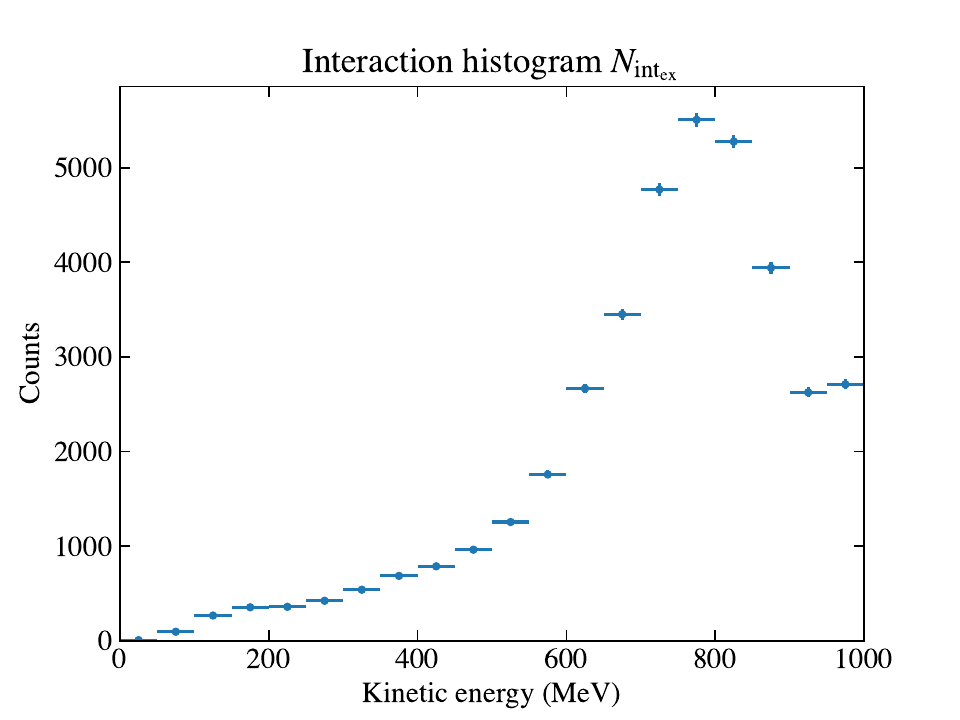}\put(-185,130){(d)}
    \caption{Energy histograms derived from the simulation sample: (\textbf{a}) $N_{\rm initial}(E)$, (\textbf{b}) $N_{\rm end}(E)$, \linebreak (\textbf{c}) $N_{\rm interaction_{ex}}(E)$, and (\textbf{d}) $N_{\rm incident}(E)$. The first and last energy bins are given as overflows. The derivation of error bars is described later in Section~\ref{sec:Deriving the statistical uncertainty}.}
    \label{fig:Nhist}
\end{figure}

\begin{figure}[H]
    \centering
    \includegraphics[width=0.9\columnwidth]{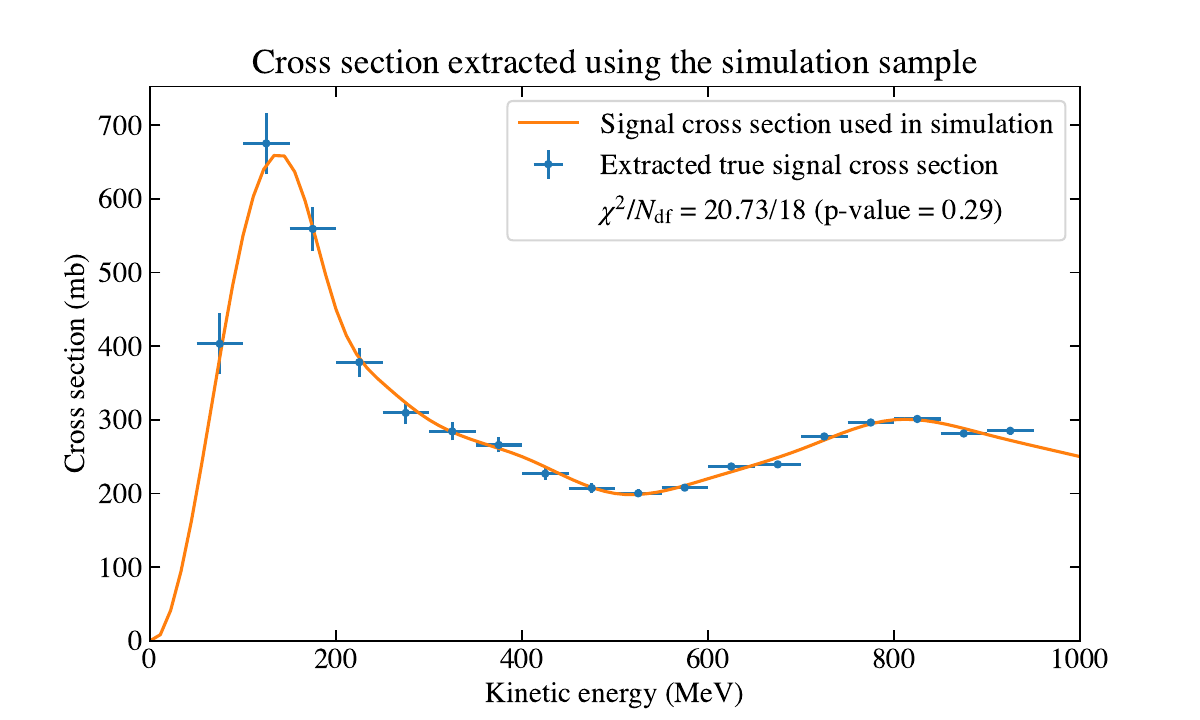}
    \caption{The cross section extracted using the true information of the simulation sample. {The right-tail \emph{p} value was calculated assuming a $\chi^2$ distribution with the number of degrees of freedom $N_{\rm df}$ being 18, which is the number of cross section bins.} The derivation of error bars is described later \mbox{in Section}~\ref{sec:Deriving the statistical uncertainty}.}
    \label{fig:trueXS}
\end{figure}

\subsection{Deriving the Statistical Uncertainty}
\label{sec:Deriving the statistical uncertainty}
In the last Section~\ref{sec:Extracting the true cross section}, we described how to extract the true cross section but not yet how to calculate its statistical uncertainty. In order to accomplish this, we considered the problem in {the multi-dimensional variable} space. This is because the three histograms $N_{\rm initial}$, $N_{\rm end}$, and $N_{\rm interaction_{ex}}$ directly derived from each event were not independent of each other. For example, for different values of $E_{\rm ini}$, the distributions of $E_{\rm end}$ are supposed to be different. In order to fully consider the correlations among them, we defined a combined variable:
\begin{equation}
    {\rm ID_{com}}={\rm ID_{ini}}+(N_{\rm bin}+1)\cdot{\rm ID_{end}}+(N_{\rm bin}+1)^2\cdot{\rm ID_{int_{ex}}}.
    \label{eqn:IDcom}
\end{equation}

Here, we assigned ${\rm ID}=0$ for events with a null value, {which is defined} in Section~\ref{sec:Slicing method}, and thus there was one more bin in addition to the number of energy bins $N_{\rm bin}$ for each of ${\rm ID_{ini}}$, ${\rm ID_{end}}$, and ${\rm ID_{int_{ex}}}$. \footnote{The null-value events need to be included rather than cut directly, because it is possible for them to be assigned a normal value in the measured space described in Section~\ref{sec:Measurement effects}, which should be handled by the response matrix.} This definition can be understood as flattening a 3D array into a 1D array. The distribution of this combined variable is shown in Figure~\ref{fig:true_IDcom}. Since we had 20 bins for all of $N_{\rm initial}$, $N_{\rm end}$, and $N_{\rm interaction_{ex}}$, there were $(20+1)^3=9261$ bins for ${\rm ID_{com}}$. The entry in each bin was derived by counting the (weighted) events independently, and thus a (weighted) Poisson error could be assigned as the statistical uncertainty for each bin content.
\begin{figure}[H]
    \centering
    \includegraphics[width=0.8\columnwidth]{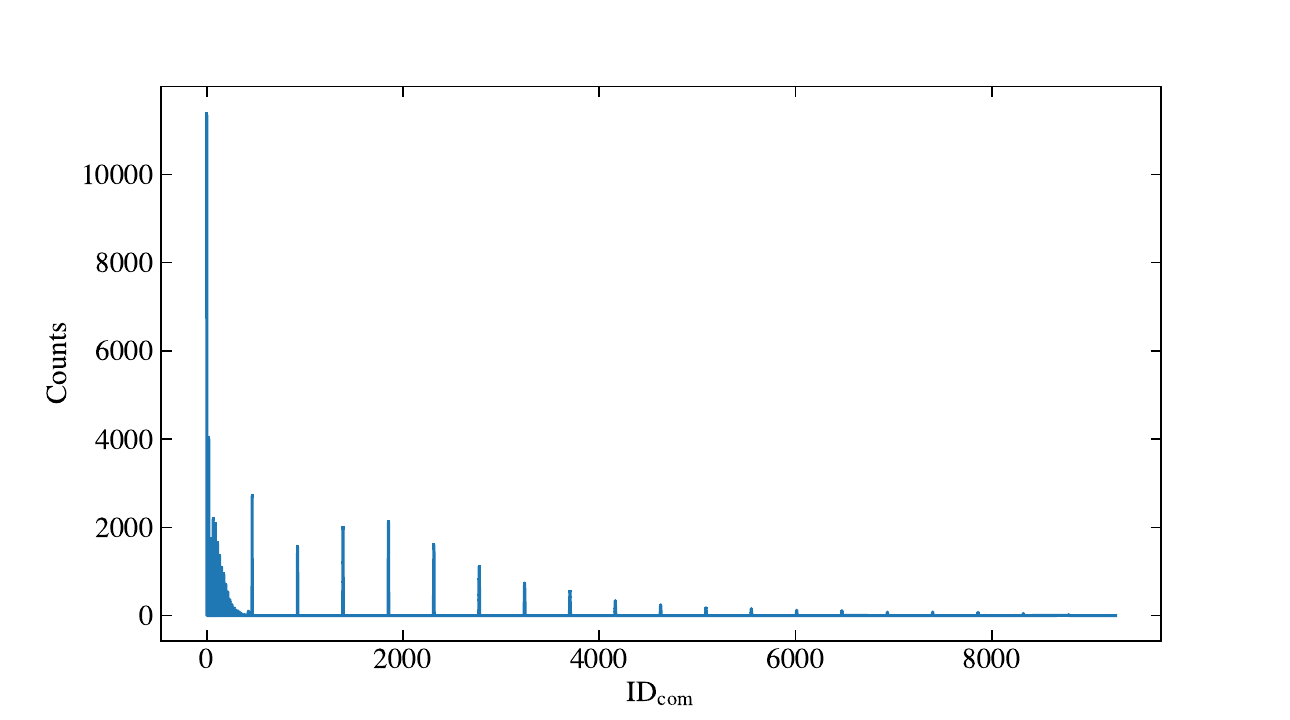}
    \caption{The
    distribution of the combined variable ${\rm ID_{com}}$ of the simulation sample.}
    \label{fig:true_IDcom}
\end{figure}

\textls[+15]{After that, we followed standard error propagation by constructing a Jacobian matrix, defined as $J=(\partial f_i/\partial x_j)_{m\times n}$, where $x$ denotes the original variable and $f$ denotes the variable it will be transformed into, and thus the covariance matrix is propagated by} \linebreak $V_{f}=J\cdot V_x\cdot J^T$. There are three steps to transition from the covariance matrix for the combined variable to the covariance matrix for the cross section, which are also described in the caption of Figure~\ref{fig:error_prop_Vtrue}\footnote{In Figure~\ref{fig:error_prop_Vtrue}a, the matrix may look to be empty because it is sparse, and the bins may be too small for readers to visualize its color. {It is still kept for consistency with the other subplots.} Similarly, this happens to Figure~\ref{fig:error_prop_Vmeas}c.}:
\begin{itemize}
    \item Firstly, the combined variable is projected back to the three axes, namely ${\rm ID_{ini}}$, ${\rm ID_{end}}$, and ${\rm ID_{int_{ex}}}$,\footnote{This can be accomplished by calculating \textls[-5]{$N_{\rm initial}=\sum_{\rm ID_{end}}\sum_{\rm ID_{int_{ex}}}({\rm ID_{ini}}$, ${\rm ID_{end}}$, ${\rm ID_{int_{ex}}})$, $N_{\rm end}=\sum_{\rm ID_{ini}}\sum_{\rm ID_{int_{ex}}}({\rm ID_{ini}}$, ${\rm ID_{end}}$, ${\rm ID_{int_{ex}}})$, and $N_{\rm interaction_{ex}}=\sum_{\rm ID_{ini}}\sum_{\rm ID_{end}}({\rm ID_{ini}}$, ${\rm ID_{end}}$,}\linebreak ${\rm ID_{int_{ex}}})$, where $N$ denotes the bin content for the corresponding ${\rm ID}$} and the covariance matrix for $({\rm ID_{ini}}; {\rm ID_{end}}; {\rm ID_{int_{ex}}})$ is derived.
    \item Secondly, the null value bins with ${\rm ID}=0$ for the three variables are ignored. $N_{\rm inc}$, calculated by Equation~\eqref{eqn:Ninc}, replaces $N_{\rm ini}$, and the covariance matrix for $(N_{\rm inc}; N_{\rm end};$ $N_{\rm int_{ex}})$ is derived.
    \item \textls[-15]{Thirdly, the cross section is calculated using the three energy histograms with \mbox{Equation~\eqref{eqn:XS_energy},}} and its covariance matrix can be derived by calculating the derivatives appearing in the Jacobian matrix.
\end{itemize}
With these covariance matrices, the error bars in Figures~\ref{fig:Nhist} and~\ref{fig:trueXS} were obtained. Figure~\ref{fig:error_prop_Vtrue} shows the correlation matrices for these four stages, allowing better visualization than the covariance matrices. As can be seen in Figure~\ref{fig:error_prop_Vtrue}c, the off-diagonal blocks were not empty, which suggests there were correlations among different histograms. In Figure~\ref{fig:error_prop_Vtrue}d, the correlation matrix for $\sigma(E)$ is diagonal, which suggests that the true cross section in each bin was independent. However, this would not be the case for the measured cross section, as shown later in Section~\ref{sec:Fake data results}. Thus, considering the problem in {the multi-dimensional variable} space is necessary for a rigorous uncertainty evaluation.
\begin{figure}[H]
    \centering
    \includegraphics[width=0.48\columnwidth]{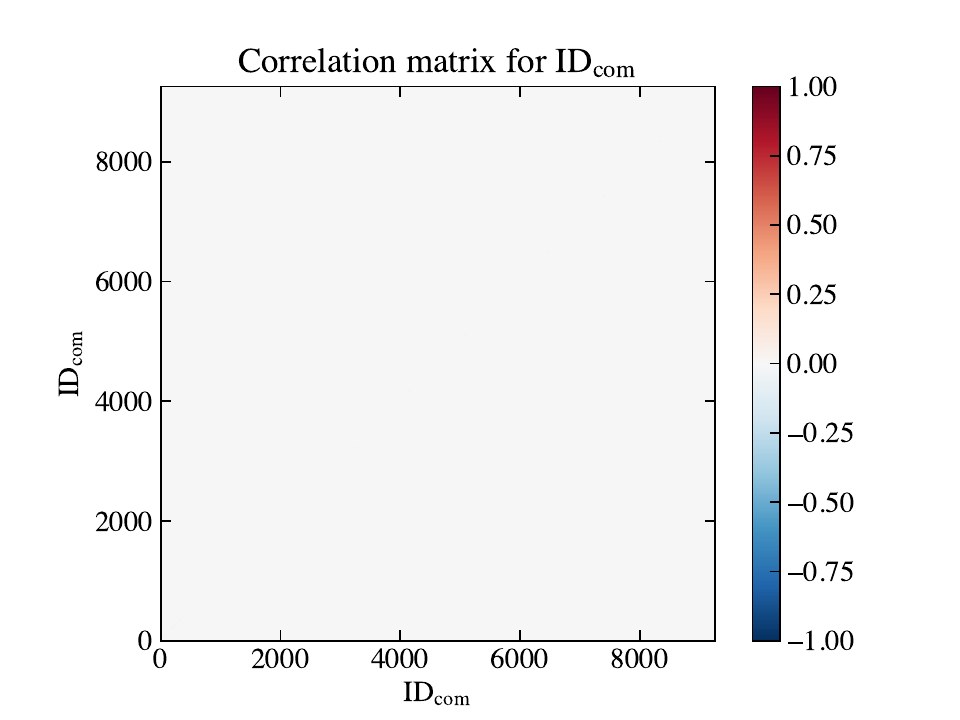}\put(-185,130){(a)}
    \includegraphics[width=0.48\columnwidth]{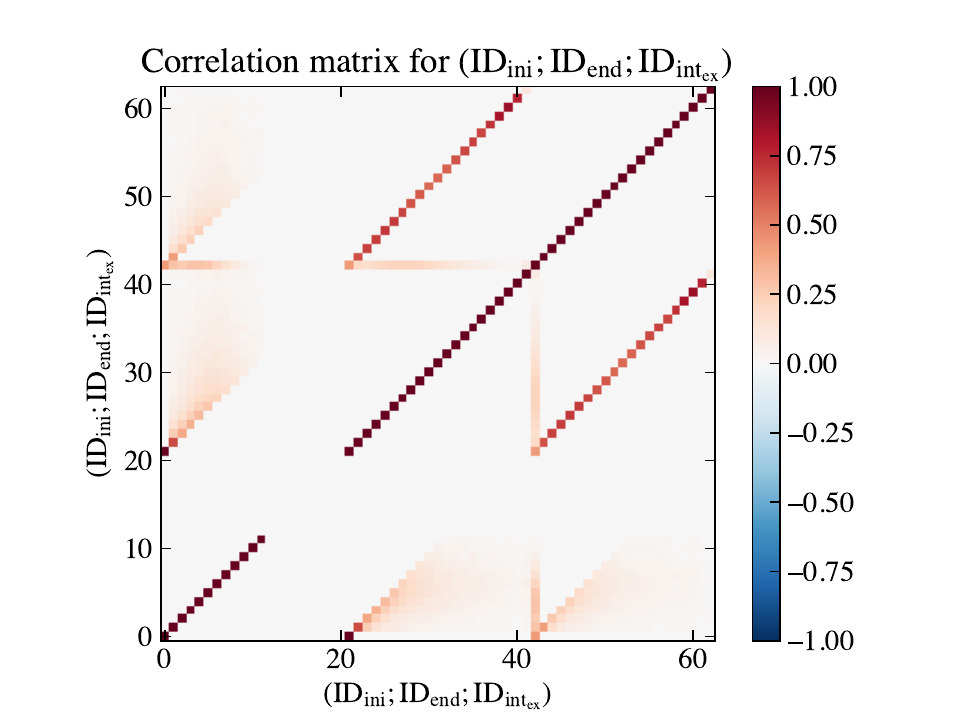}\put(-185,130){(b)}
    
    \includegraphics[width=0.48\columnwidth]{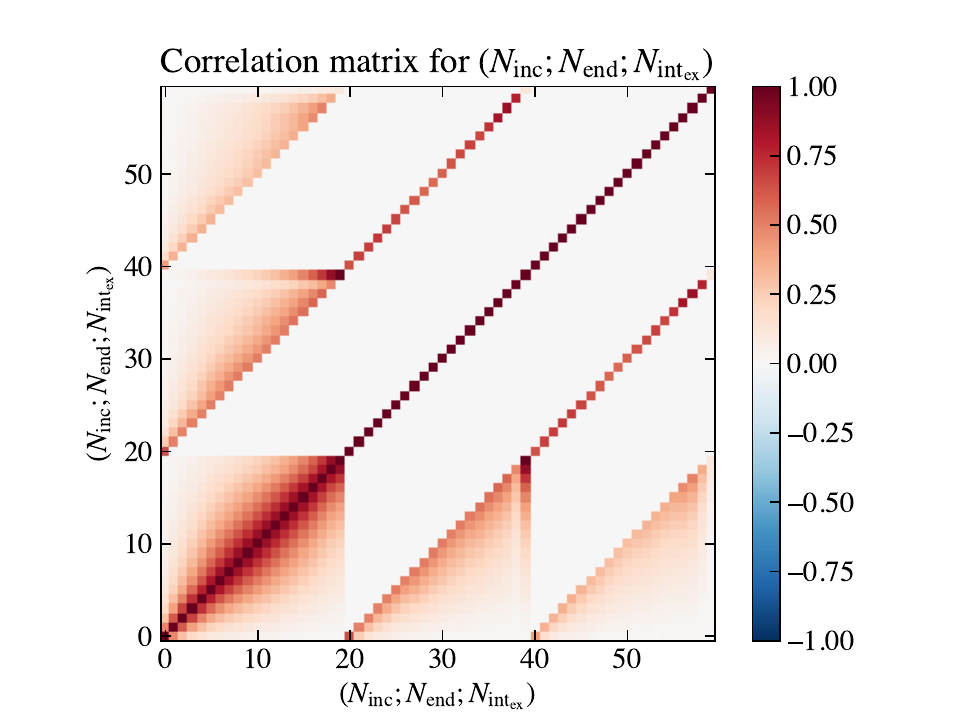}\put(-185,130){(c)}
    \includegraphics[width=0.48\columnwidth]{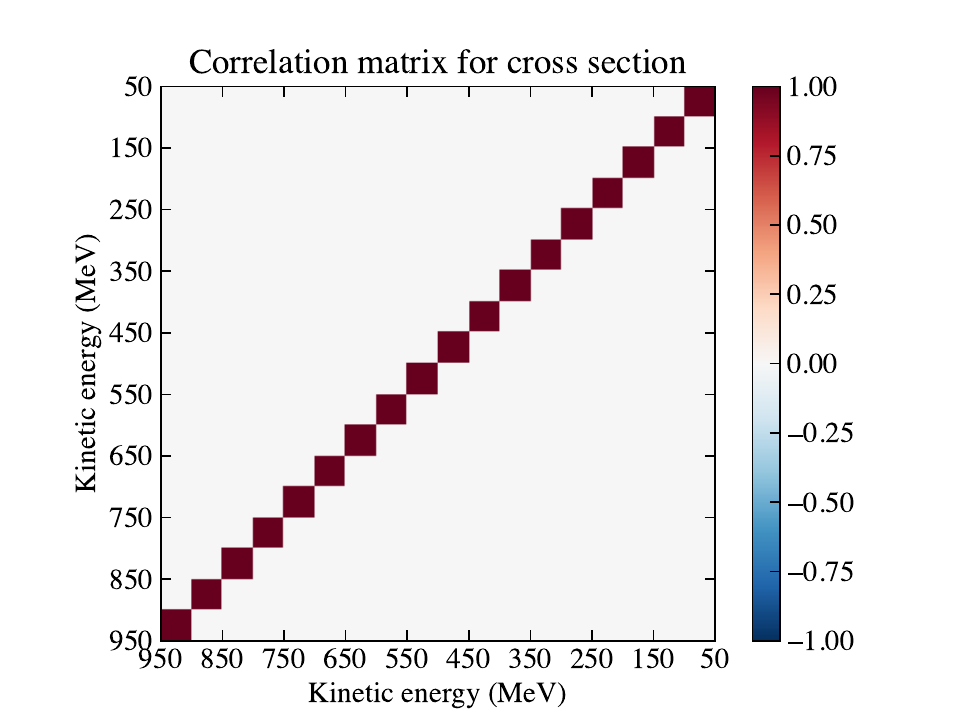}\put(-185,130){(d)}
    \caption{(\textbf{a}) Correlation matrix for the combined variable ${\rm ID_{com}}$, which is diagonal since the entry in each bin was derived by counting independently. (\textbf{b}) Correlation matrix for $({\rm ID_{ini}}; {\rm ID_{end}}; {\rm ID_{int_{ex}}})$, where the first block (bin indices 0--20) corresponds to ${\rm ID_{ini}}$, the second block (bin indices 21--41) corresponds to ${\rm ID_{end}}$, and the third block (bin indices 42--62) corresponds to ${\rm ID_{int_{ex}}}$. (\textbf{c}) Correlation matrix for $(N_{\rm inc}; N_{\rm end}; N_{\rm int_{ex}})$, where the first block (bin indices 0--19) corresponds to $N_{\rm inc}$, the second block (bin indices 20--39) corresponds to $N_{\rm end}$, and the third block (bin indices 40--59) corresponds to $N_{\rm int_{ex}}$. (\textbf{d}) Correlation matrix for the extracted cross section $\sigma(E)$, which has 18 bins on each axis without the underflow and overflow bins.}
    \label{fig:error_prop_Vtrue}
\end{figure}

\subsection{Measurement Effects}
\label{sec:Measurement effects}
In practice, the true values of $E_{\rm ini}$ and $E_{\rm end}$ and the type of interaction are unknown, but they need to be measured, which may result in some measurement effects, including the detector resolution and inefficiency when measuring the energy, as well as misidentification of the type of interaction. $E_{\rm ini}$ is usually measured with external instruments outside the LArTPC, while $E_{\rm end}$ is derived from $E_{\rm ini}$ with the reconstructed track information in the LArTPC. We also relied on random variables to model these effects effectively rather than simulating events from the origins of the effects. Firstly, to simulate the selection process, a score was generated for each event. Based on whether the score was larger or smaller than a threshold, the event was kept or rejected. The score was generated following a Gaussian distribution, whose mean parameter depended on the true parameters of the event, in order for the efficiency to be non-uniform as a more general case.\footnote{In more detail, the mean of the Gaussian distribution used for the simulation was a linear function of $\log{\left(E_{\rm ini}^{\rm true}-E_{\rm end}^{\rm true}\right)}$, and a constant threshold needed to be reached for the event to be selected. This mimics the actual scenario where the selection efficiency is smaller for too short of tracks and larger if the tracks are long enough.} In all, {40,917 out of 100,000 events in the simulation sample passed selection.} Secondly, for all events that passed selection, two random variables following different Gaussian distributions were generated for each event, denoted as $E_1$ and $E_2$. $E_1$ was used to imitate the resolution for measuring $E_{\rm ini}$, while $E_2$ accounted for the resolution effects involved with energy deposition of the reconstructed track. Therefore, we had $E_{\rm ini}^{\rm meas}=E_{\rm ini}^{\rm true}+E_1$ and $E_{\rm end}^{\rm meas}=E_{\rm end}^{\rm true}+E_1+E_2$\footnote{$E_{\rm end}^{\rm meas}$ was derived from the measurement of $E_{\rm ini}^{\rm meas}$, and thus $E_1$ was inherited}. Finally, in order to simulate the misidentification among {the three types} of interactions, or to say the event fates, which were ``no interaction'', ``signal interaction'', and ``other interactions'', a $3\times 3$ confusion matrix was defined, where each element indicated the possibility of a true fate being recognized as a measured fate. Random numbers were used in order to decide the measured fate of each event according to the defined confusion matrix. As a result, the simulated measurement effects\footnote{In this simulation, the resolutions and the confusion matrix were defined as the same for all events, which were independent of their energy. The work in~\cite{hadron-Ar_XS} can be referred to for more detail.} can be seen in Figure~\ref{fig:measuring_effects}.
\begin{figure}[H]
    \centering
    \includegraphics[width=0.48\columnwidth]{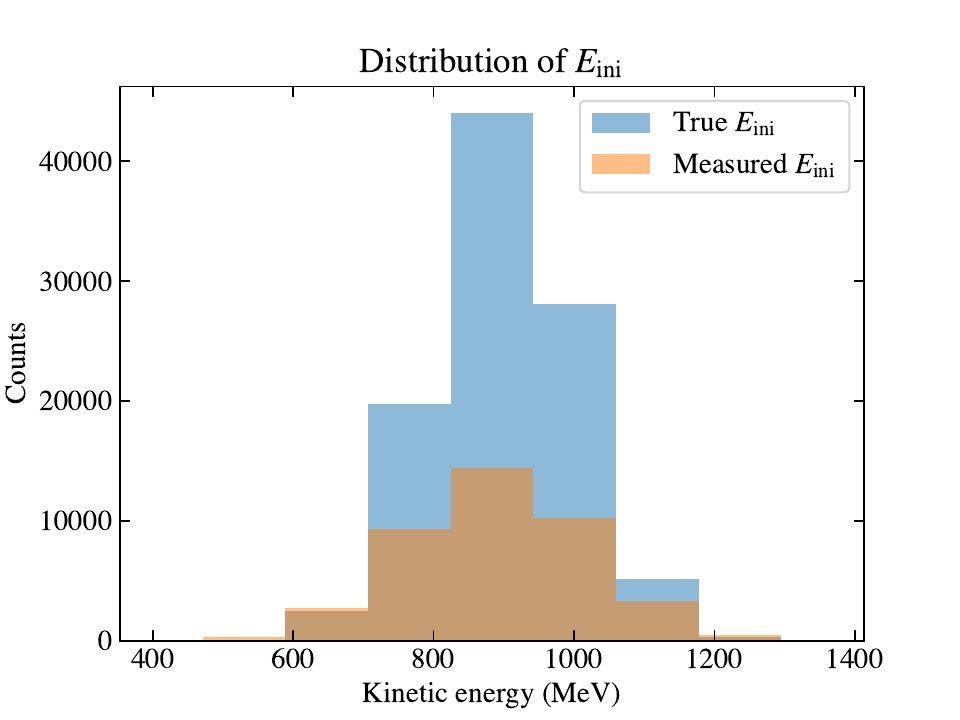}\put(-185,130){(a)}
    \includegraphics[width=0.48\columnwidth]{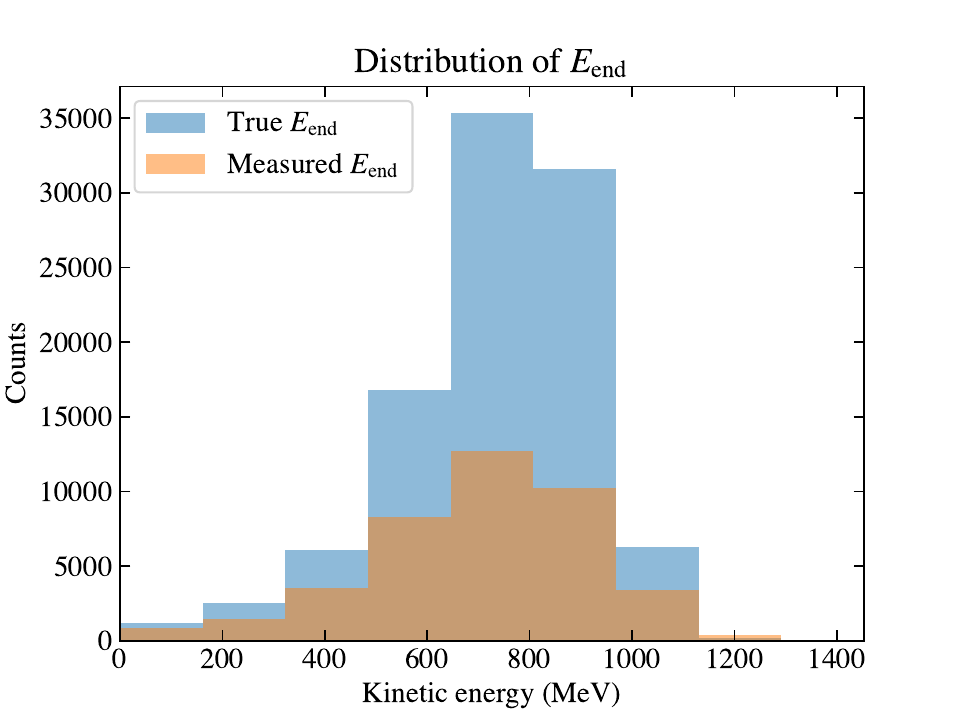}\put(-185,130){(b)}
    
    \includegraphics[width=0.48\columnwidth]{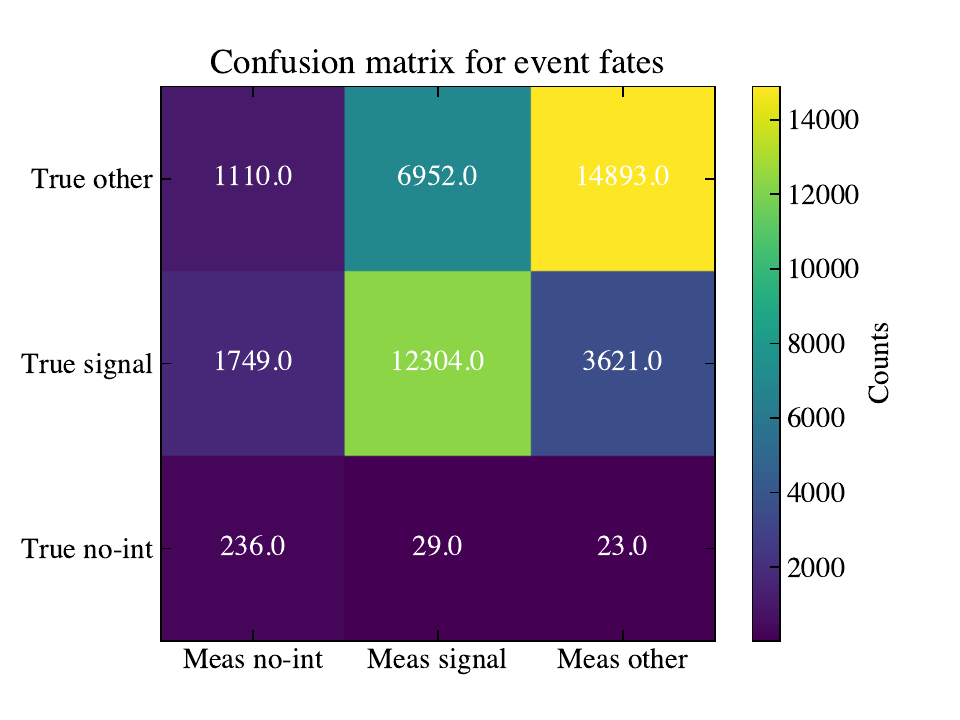}\put(-185,130){(c)}
    \caption{(\textbf{a}) The distribution of $E_{\rm ini}^{\rm meas}$ (orange histogram) and $E_{\rm ini}^{\rm true}$ (blue histogram) for the simulation sample. (\textbf{b}) The distribution of $E_{\rm end}^{\rm meas}$ (orange histogram) and $E_{\rm end}^{\rm true}$ (blue histogram) for the simulation sample. (\textbf{c}) The resulting confusion matrix for the event fates of the simulation sample. The horizontal axis indicates the measured fates, and the vertical axis indicates the true fates. The color bar indicates the (weighted) event counts in each bin, which add up to be the event counts passing selection.}
    \label{fig:measuring_effects}
\end{figure}

Therefore, for each event in the simulation sample passing selection, it had three true properties and three measured properties. With these properties, we were able to determine its index for the combined variable ${\rm ID_{com}}$ in both the true space and the measured space, and thus we could use the simulation sample to model the response matrix as well as the efficiency plot for ${\rm ID_{com}}$, which would later be the inputs for unfolding~\cite{Cowan:2002in} as well as efficiency correction. Figure~\ref{fig:meas_IDcom} shows the distribution of the measured combined variable ${\rm ID_{com}^{\rm meas}}$, which was also calculated by Equation~\eqref{eqn:IDcom} but with the measured values. 
\begin{figure}[H]
    \centering
    \includegraphics[width=0.8\columnwidth]{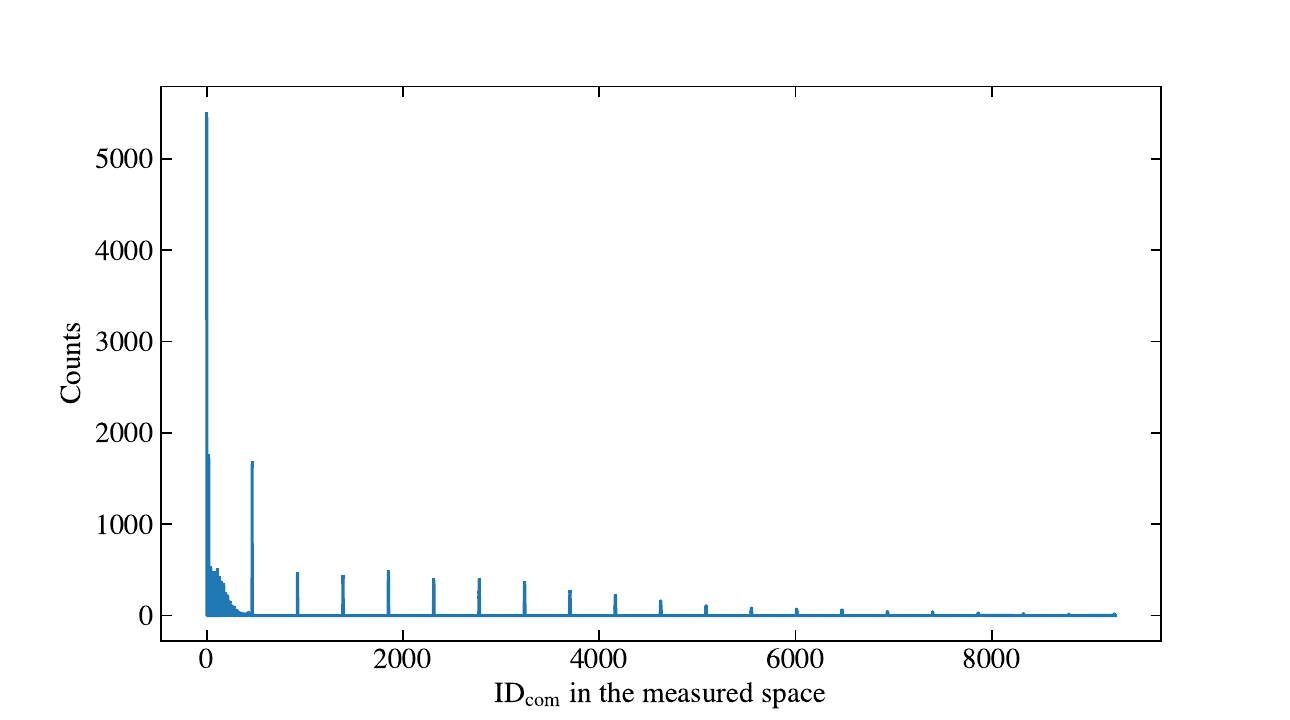}
    \caption{The distribution of the measured combined variable ${\rm ID_{com}^{\rm meas}}$ of the simulation sample. }
    \label{fig:meas_IDcom}
\end{figure}

For the real data, the true information was not available, and thus we used unfolding together with the efficiency correction to transform from the measured histogram to the estimated true histogram, which is referred to as the unfolded histogram. However, the number of bins for ${\rm ID_{com}}$ to the order of $N_{\rm bins}^3$ could reach over a thousand. To unfold a histogram with such a large number of bins can be quite time-consuming. Fortunately, in our case, despite the large number of bins, the histogram was usually sparse, with many empty bins. \footnote{For example, out of 9261 bins in total, for either the ${\rm ID_{com}^{\rm true}}$ histogram in Figure~\ref{fig:true_IDcom} or the ${\rm ID_{com}^{\rm meas}}$ histogram in Figure~\ref{fig:meas_IDcom}, only 311 bins and 350 bins were non-empty, respectively.} Therefore, we deleted these empty bins in the ${\rm ID_{com}^{true}}$ histogram and the ${\rm ID_{com}^{meas}}$ histogram separately, re-indexed the remaining bins, and denoted the new index as ${\rm ID_{rem}}$. A map was created between ${\rm ID_{com}^{true}}$ and ${\rm ID_{rem}^{true}}$, and another map was created between ${\rm ID_{com}^{meas}}$ and ${\rm ID_{rem}^{meas}}$. After that, we built the response matrix as well as the efficiency plot of ${\rm ID_{rem}}$, as shown in Figure~\ref{fig:response}.
\begin{figure}[H]
    \centering
    \includegraphics[width=0.43\columnwidth]{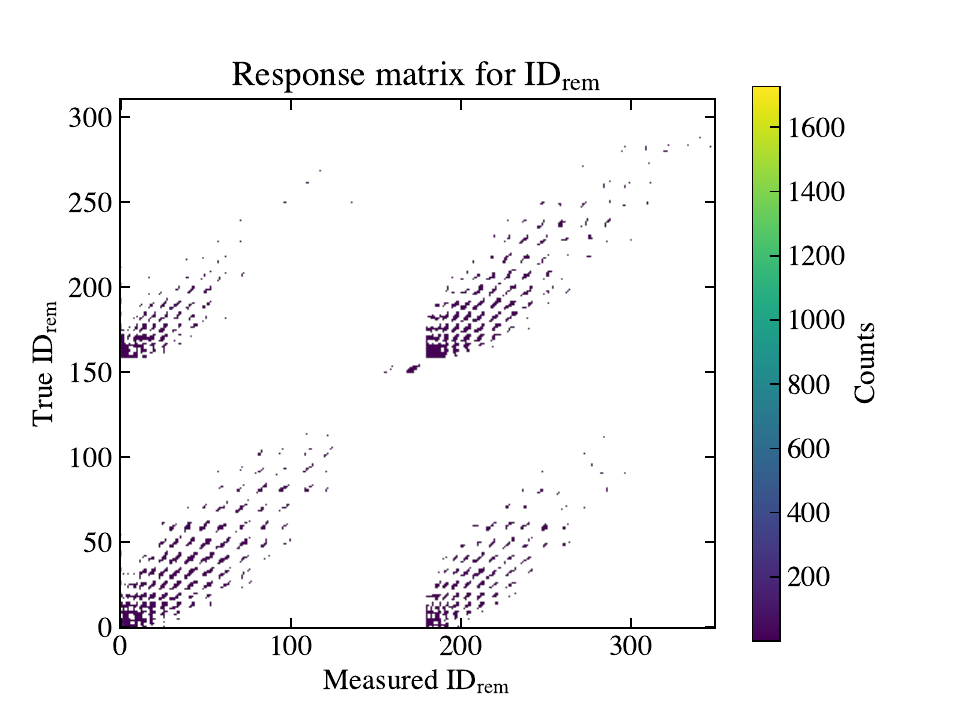}\put(-160,115){(a)}
    \includegraphics[width=0.57\columnwidth]{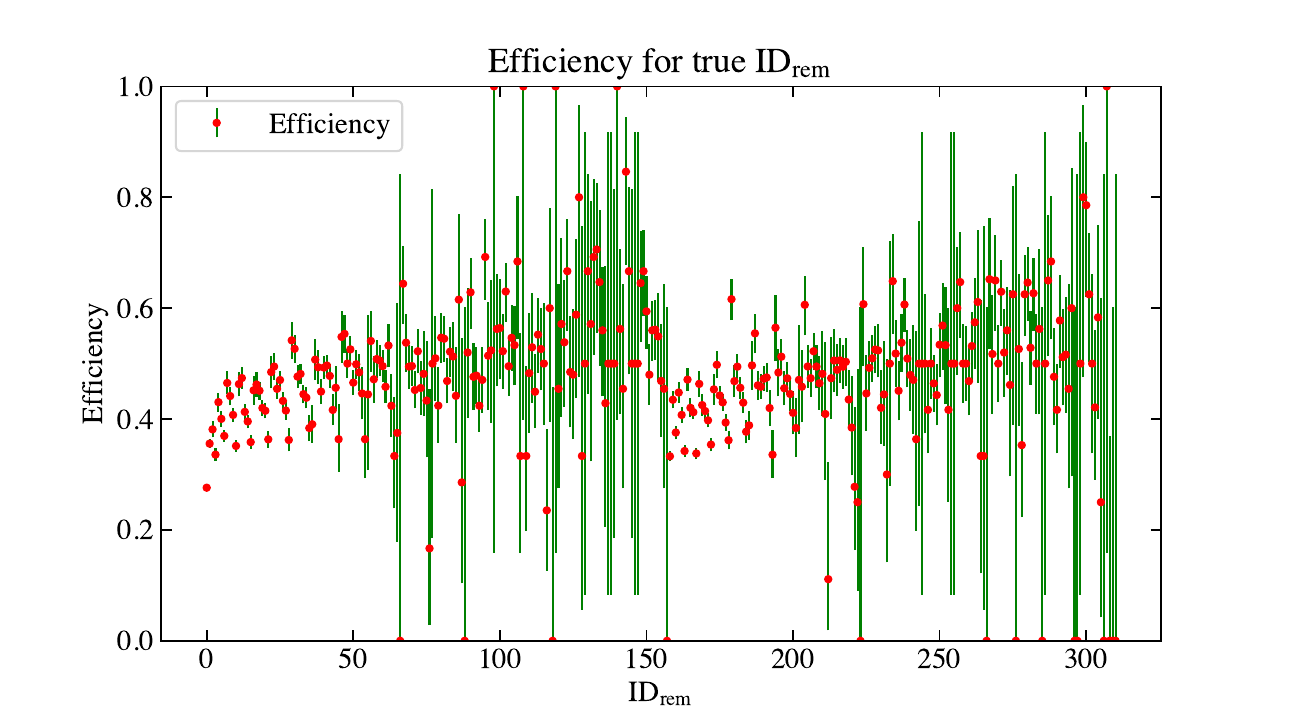}\put(-215,115){(b)}
    \caption{(\textbf{a}) The response matrix modeled using the simulation sample. The horizontal axis is ${\rm ID_{rem}^{meas}}$, and the vertical axis is the true ${\rm ID_{rem}^{true}}$. The color bar indicates the (weighted) event counts for each bin. (\textbf{b}) The efficiency for each ${\rm ID_{rem}^{true}}$ bin. The uncertainty for efficiency was calculated according to the Clopper--Pearson method~\cite{10.1093/biomet/26.4.404}.}
    \label{fig:response}
\end{figure}

By denoting the response matrix as $R_{ij}=P({\rm ID_{rem}^{meas}}=j|{\rm ID_{rem}^{true}}=i)$ and the efficiency as $\epsilon_i=P({\rm events\,\,with\,\,{ID_{rem}^{true}}=i\,\,being\,\,selected})$, we had ${\rm ID_{rem}^{meas;sim}}=R\cdot(\epsilon\cdot{\rm ID_{rem}^{true;sim}})$, where ${\rm ID_{rem}^{true}}$ held for the simulation sample. For a data sample, we first used the map for the measured histogram to derive ${\rm ID_{rem}^{meas}}$ from ${\rm ID_{com}^{meas}}$. Then, we relied on an unfolding algorithm of choice to estimate the unfolding matrix, denoted as $\widetilde{R}$,\footnote{It might be natural to think of $\widetilde{R}$ as the direct inverse of $R$, but it has been proven to be problematic to use. This has been described in many references about unfolding, such as~\cite{Cowan:2002in}.} and thus the unfolded ${\rm ID_{rem}}$ histogram for the data was ${\rm ID_{rem}^{unfd;data}}=(\widetilde{R}\cdot{\rm ID_{rem}^{meas;data}})\cdot\epsilon^{-1}$.\footnote{If $\epsilon$ in bin $i$ is zero, then the value in the bin ${\rm ID_{rem}^{unfd;data}}=i$ can be estimated using the simulation sample normalized to the data sample directly, because the zero efficiency is usually due to low statistics and will not change the final result significantly. However, the uncertainty associated with this can be evaluated by fluctuating these bin entries.} Finally, the map for the true histogram was used to transform ${\rm ID_{rem}^{true}}$ back into ${\rm ID_{com}^{true}}$.\footnote{It is possible that a bin for ${\rm ID_{com}}$ is empty in the simulation sample but not empty in the data sample, especially for bins with low statistics. In this case, we can add these non-empty bins of data to the map as well. It is not necessary for the map to be the same for all data samples.}

\subsection{Fake Data Results}
A sample of 10,000 events was generated using the same procedure as the simulation sample, but its true information was not used in order to mimic the real data. After selection, we had 4,190 events in this fake data sample. {The selection rate was similar to the simulation sample. }Figure~\ref{fig:error_prop_Vmeas} shows the correlation matrices involved in the error propagation from the measured ${\rm ID_{rem}}$ histograms to the final cross section results. Compared with Figure~\ref{fig:error_prop_Vtrue}, when extracting the true cross section, there were two extra steps, which were unfolding and mapping back to ${\rm ID_{rem}}$.
\label{sec:Fake data results}
\begin{figure}[H]
    \centering
    \includegraphics[width=0.48\columnwidth]{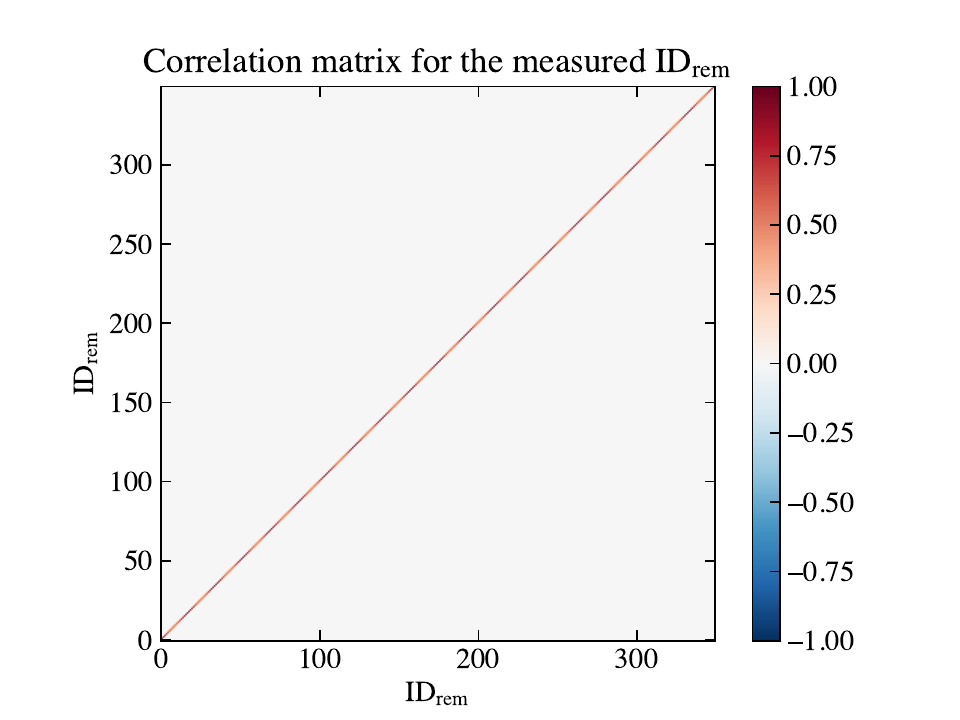}\put(-185,130){(a)}
    \includegraphics[width=0.48\columnwidth]{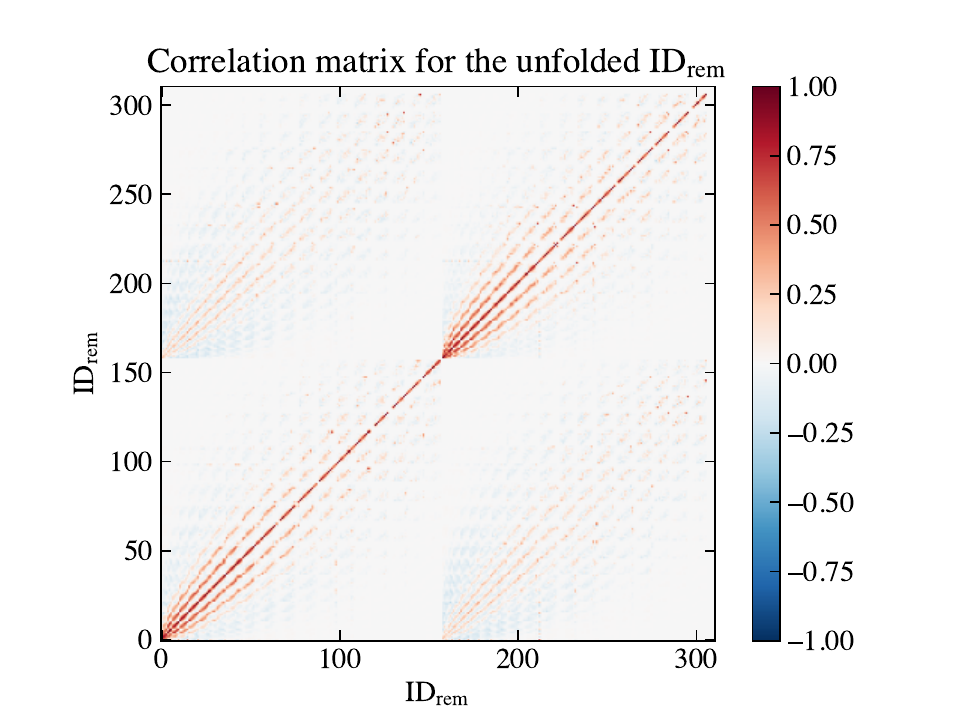}\put(-180,130){(b)}
    
    \includegraphics[width=0.48\columnwidth]{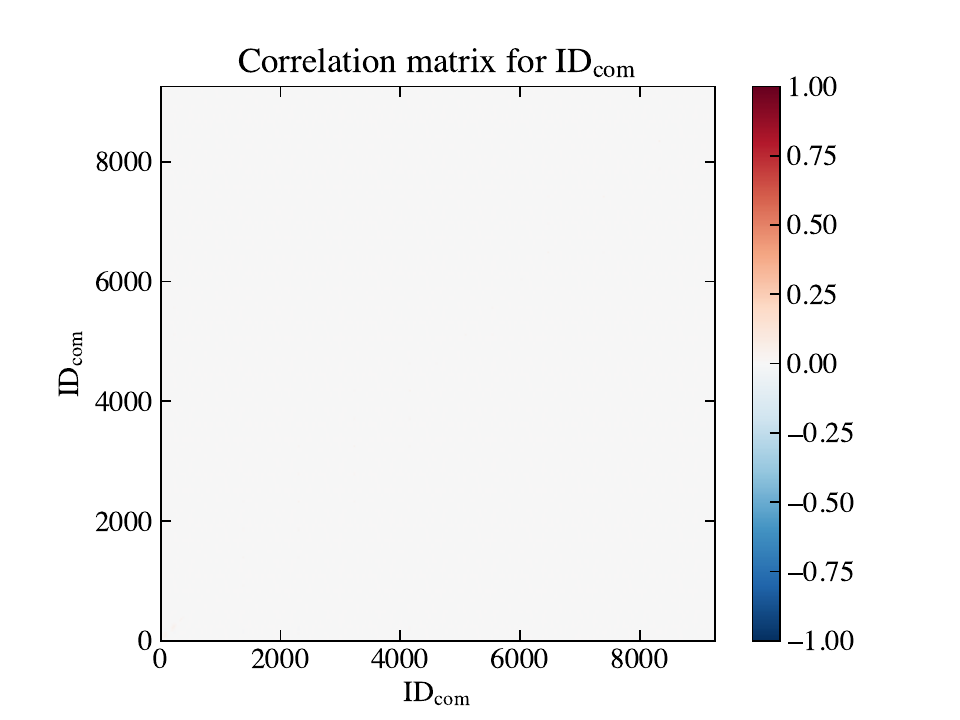}\put(-185,130){(c)}
    \includegraphics[width=0.48\columnwidth]{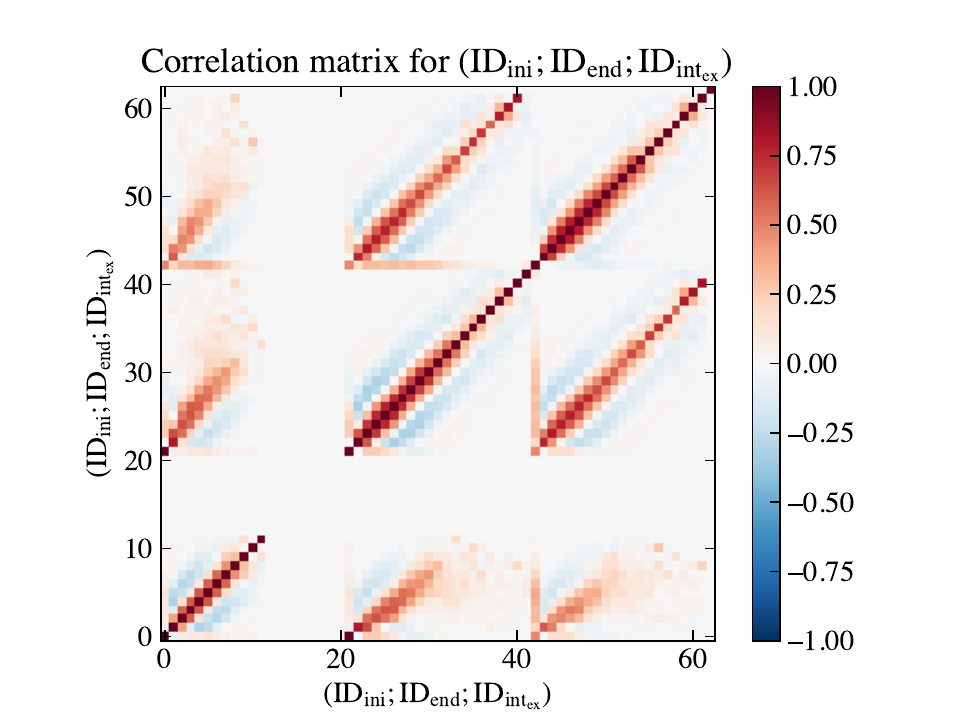}\put(-180,130){(d)}

    \includegraphics[width=0.48\columnwidth]{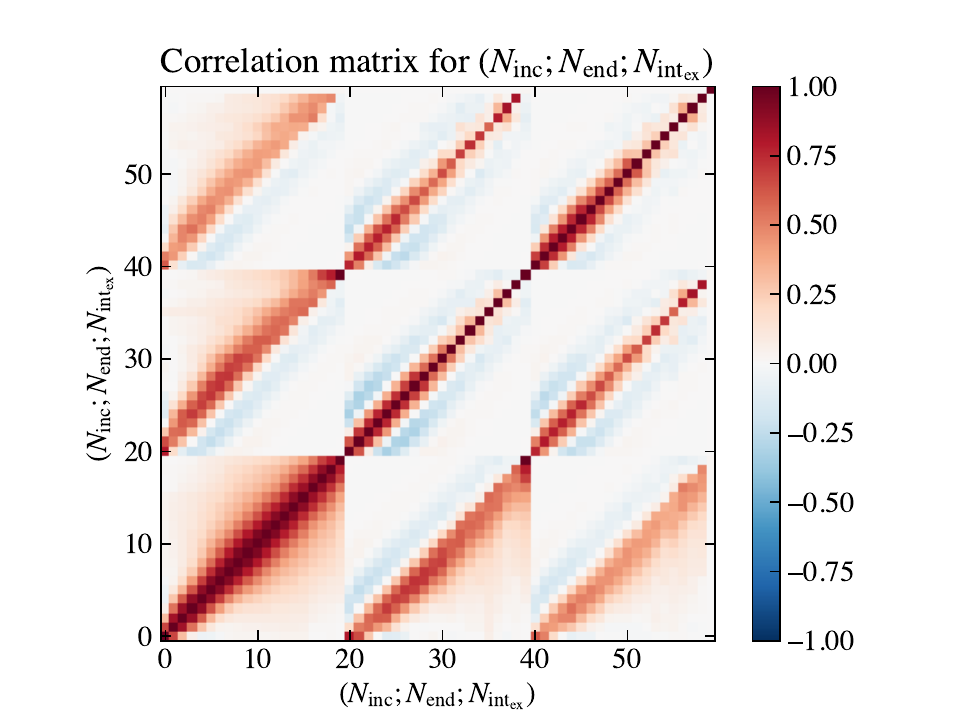}\put(-185,130){(e)}
    \includegraphics[width=0.48\columnwidth]{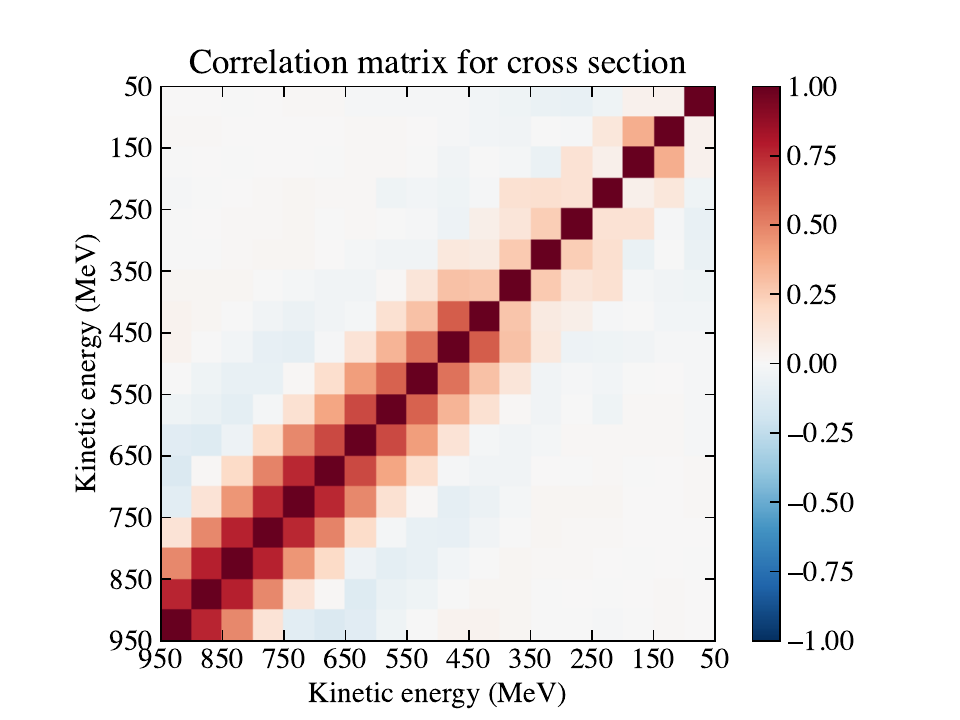}\put(-180,130){(f)}
    \caption{Correlation matrices for (\textbf{a}) ${\rm ID_{rem}^{meas}}$, (\textbf{b}) ${\rm ID_{rem}^{unfd}}$, (\textbf{c}) ${\rm ID_{com}}$, (\textbf{d}) $({\rm ID_{ini}}; {\rm ID_{end}}; {\rm ID_{int_{ex}}})$, \linebreak (\textbf{e}) $(N_{\rm inc}; N_{\rm end}; N_{\rm int_{ex}})$, and (\textbf{f}) the measured cross section $\sigma(E)$ for the fake data sample.}
    \label{fig:error_prop_Vmeas}
\end{figure}

In Figure~\ref{fig:error_prop_Vmeas}a, the correlation matrix for ${\rm ID_{rem}^{meas}}$ is diagonal because in each bin, the events were counted independently. Figure~\ref{fig:error_prop_Vmeas}b is the correlation matrix for ${\rm ID_{rem}^{unfd}}$, which was derived from (a) using the unfolding algorithm of choice. We used the Python version of RooUnfoldBayes~\cite{Brenner:2019lmf} as the unfolding algorithm, which employs the d'Agostini method~\cite{DAgostini:1994fjx}. The unfolding algorithm was treated as a black box in this paper, where we input the measured ${\rm ID_{rem}}$ histograms as well as its covariance matrix and obtained the output, including the unfolded ${\rm ID_{rem}}$ histograms as well as its covariance matrix.\footnote{However, because the ${\rm ID_{rem}}$ histogram we unfolded was not a smooth physical spectrum, the unfolding algorithms that tried to regularize the unfolded result by smoothing may thus not be applicable.} We proceeded with the number of iterations being four, which was the only parameter in d'Agostini's method for unfolding, and it was not optimized in this work. After that, ${\rm ID_{rem}}$ was converted back into ${\rm ID_{com}}$ under the map created using the measured histograms, and then the following steps were the same as those for the error propagation of the true cross section, as described in Section~\ref{sec:Deriving the statistical uncertainty}. Figure~\ref{fig:measXS} shows the cross section measured using the fake data sample, whose correlation matrix is shown in Figure~\ref{fig:error_prop_Vmeas}f. The correlation matrix is not diagonal, which means the measured cross section in each energy bin was correlated, and thus for the final cross section result, we needed to present both the central values as well as their covariance matrix. We could also calculate $\chi^2$ of the measured cross section against the simulation curve with Equation~\eqref{eqn:chi2}, whose value is shown in the legend of Figure~\ref{fig:measXS}. The error bars in the lower-energy bins tended to be larger, mostly because the statistics {were} smaller, since most hadrons interact before they reach these low energies. This trend can also be seen in the true cross section in Figure~\ref{fig:trueXS} for the same reason.
\begin{figure}[H]
    \centering
    \includegraphics[width=0.9\columnwidth]{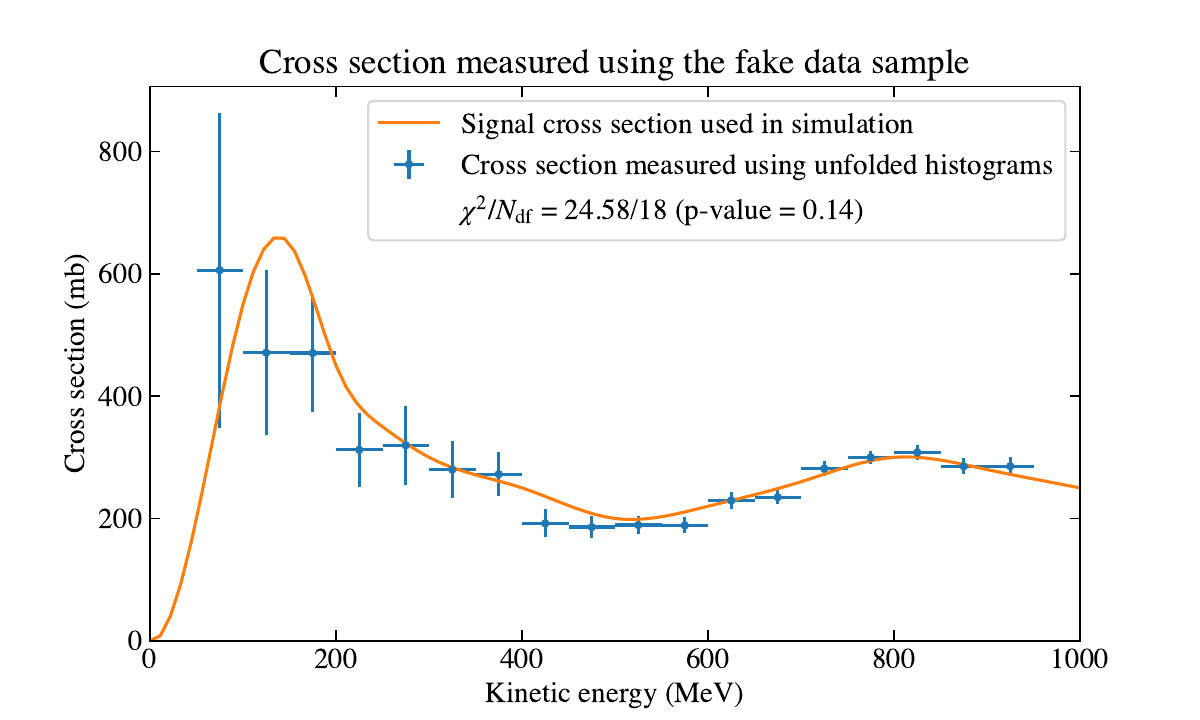}
    \caption{The cross section measured using the unfolded histograms of the fake data sample. {The right-tail p value was calculated assuming a $\chi^2$ distribution with the number of degrees of freedom $N_{\rm df}$ being 18, which is the number of cross section bins.}}
    \label{fig:measXS}
\end{figure}

\section{Discussions and Summary}
\label{sec:Discussions and summary}
In the previous section, we described how to extract the true cross section of the simulation sample as well as how to measure the cross section of a data sample. Here, $\chi^2$ was calculated in both cases, which was used to quantify the consistency against the simulation curve. In order to further test the results, we performed toy studies. Four hundred simulation samples, referred to as toys, each with a sample of 10,000 events, were generated in the same way as what was described in Section~\ref{sec:Procedures and results}. The true cross section as well as its covariance matrix was calculated in each toy simulation sample. For each cross section bin, we calculated the pull value of each toy, which is defined as
\begin{equation}
    {\rm pull}=\frac{\sigma(E)-\sigma_{\rm curve}(E)}{\sqrt{V_\sigma(E, E)}},
\end{equation}
where $\sqrt{V_\sigma(E, E)}$ is the uncertainty for $\sigma(E)$. In each bin, the pull values were expected to follow a normal distribution. Figure~\ref{fig:toy_validation_true}a shows the test results, where a Gaussian distribution was fitted onto the pull histograms in each cross section bin, as shown by the blue error bars. By visually comparison with the reference lines, we can see that they were generally consistent with the expectation that each of them centered at zero and had a bar length of one, corresponding to the two parameters in the Gaussian fit. As another test, which took into account the covariance among cross section bins, we show in Figure~\ref{fig:toy_validation_true}b the histogram of $\chi^2$, calculated according to Equation~\eqref{eqn:chi2}, for each toy. A $\chi^2$ distribution was fitted onto the histogram, whose degree of freedom $N_{\rm df}$, as shown in the legend, was consistent with the expectation, with 18 being the number of cross section bins. These tests served as a validation of the slicing method. They also suggest that given the current statistics of the events for each toy, the bias caused by the approximations of the method\footnote{For example, we considered the cross section calculated to be at the middle point in each energy bin, and we evaluated $dE/dx$ at the middle energy value as well. Further corrections are needed if the statistical uncertainty becomes smaller when the sample size is much larger.} \mbox{was insignificant.}
\begin{figure}[H]
    \centering
    \includegraphics[width=6.8cm]{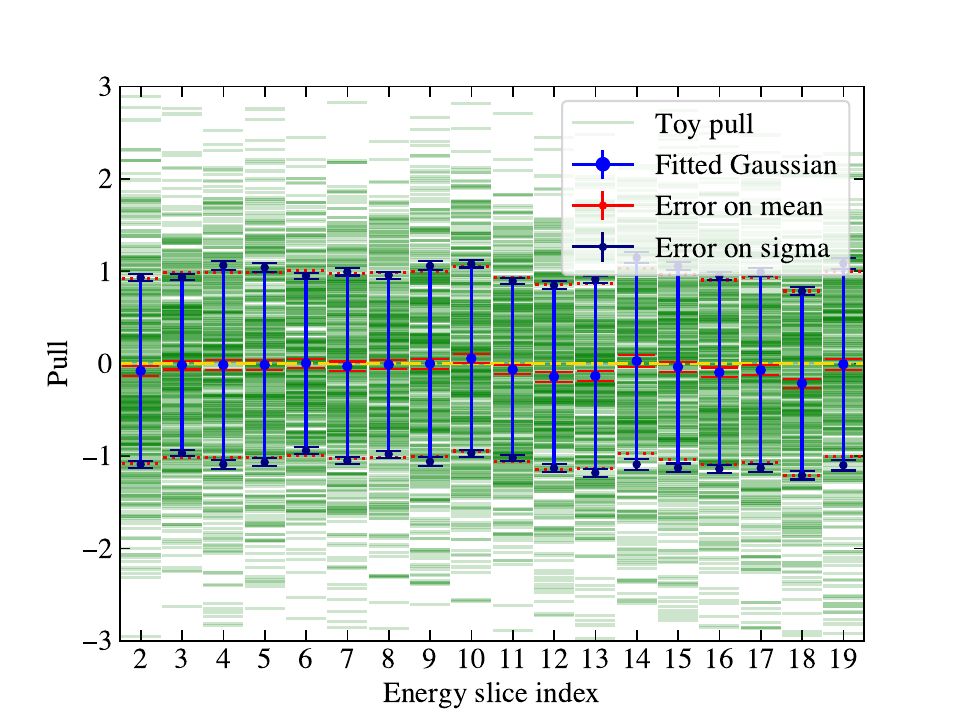}\put(-190,130){(a)}
    \includegraphics[width=6.8cm]{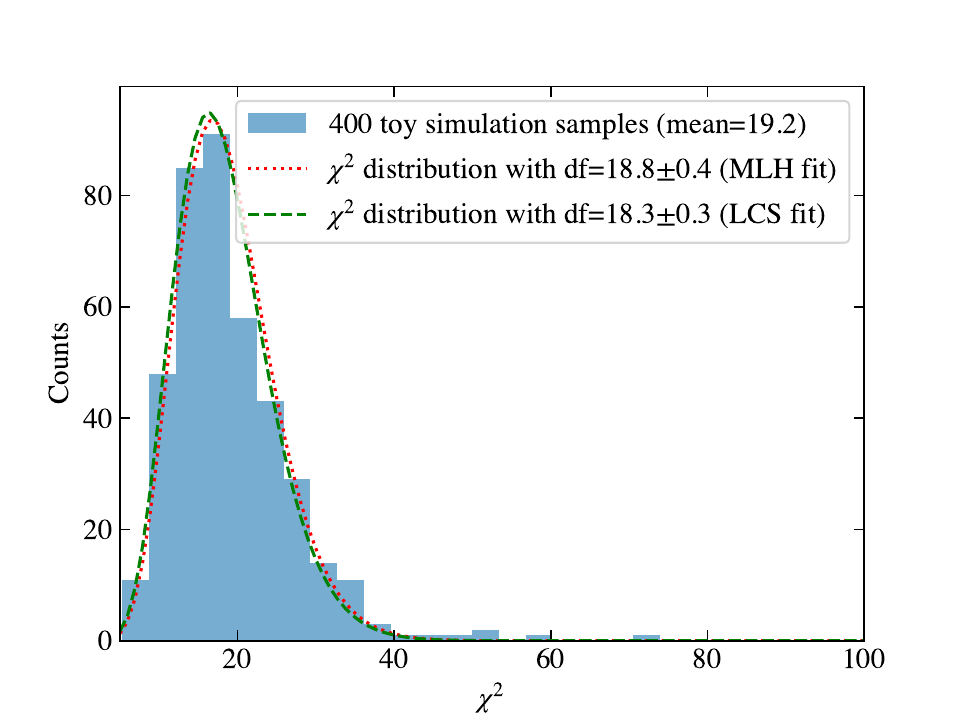}\put(-190,130){(b)}
    \caption{Toy studies using the extracted true cross sections of 400 toy simulation samples. (\textbf{a}) The pull value test results. The horizontal axis is the energy slice index ${\rm ID}$, where ${\rm ID}=2$ corresponds to an energy bin of $[900, 950]$ MeV and ${\rm ID}=19$ corresponds to an energy bin of $[50, 100]$ MeV. The vertical axis is the pull value. The green lines are the pull values of each toy in each ${\rm ID}$ bin. The blue point and its error bars indicate $\mu$ and $\sigma$ of the Gaussian fit in each ${\rm ID}$ bin. The red lines sandwiching the blue point indicate the fit error of $\mu$ in each ${\rm ID}$ bin, which can be visually compared to the dashed orange line for its consistency with 0. The dark blue lines sandwiching the end points of the blue error bars indicate the fit error of $\sigma$ in each ${\rm ID}$ bin, which can be visually compared to the dotted red line for its consistency with 1. (\textbf{b}) The histogram of $\chi^2$ against the simulation curve. Fitted $\chi^2$ distributions using both the maximum likelihood (MLH) fit and the least chi-square (LCS) fit are overlaid, with the result given in the legend.}
    \label{fig:toy_validation_true}
\end{figure}

Similarly, we generated 400 toy fake data samples, each with a sample of 10,000 events before selection, in order to study the performance of the procedures to measure the cross section. The 400 toy simulation samples used above were combined into a total of 4,000,000 events in order to model the response matrix as well as the efficiency plot for each toy fake data sample, and thus we could ignore the statistical uncertainty of the simulation sample. The cross section was measured for each toy fake data sample, and we could also derive the pull distributions in each cross section bin as well as the histogram of $\chi^2$, as shown in Figure~\ref{fig:toy_validation_meas}. In subplot (a), we can see that the lengths of the blue error bars were generally consistent with one, but some of their central points showed a small bias away from zero. This bias is considered the unfolding error. The general unfolding result effectively applies a re-smearing matrix on the true information~\cite{Tang:2017rob}. {Treating the re-smeared truth as the truth introduces an unfolding error, and thus publishing the re-smearing matrix is suggested in order for others to consider this error when comparing the results.} In addition, the unfolding error tends to be smaller when the regularization becomes weaker with a greater number of iterations. {Since we did not include this bias, the derived $\chi^2$ was supposed to be larger, whose distribution is shown in} Figure~\ref{fig:toy_validation_meas}b.

In the fake data toy study, the simulation sample used to model the response matrix and the efficiency plot was consistent with the toy fake data samples because they were generated in the same way. When it comes to real data, we need to consider the uncertainties caused by the differences between the data and simulation, which can be estimated by fluctuating the relevant parameters of the simulation sample. Additional model validation {procedures are} essential to examine the compatibility between the data and simulation and ensure the differences were within the quoted simulation uncertainties.
\begin{figure}[H]
    \centering
    \includegraphics[width=6.8cm]{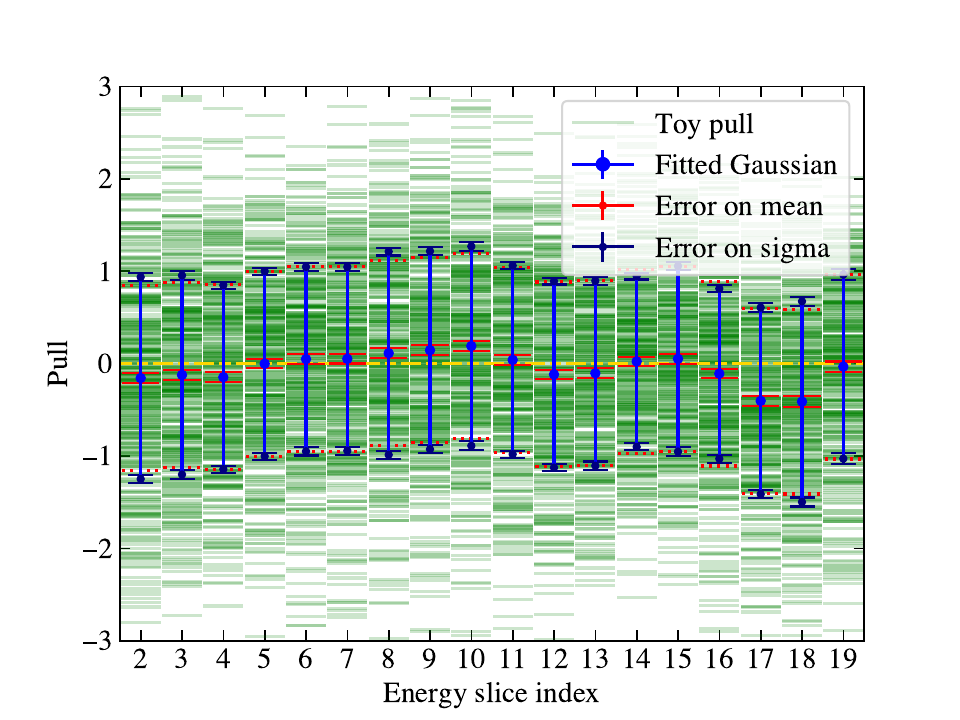}\put(-190,130){(a)}
    \includegraphics[width=6.8cm]{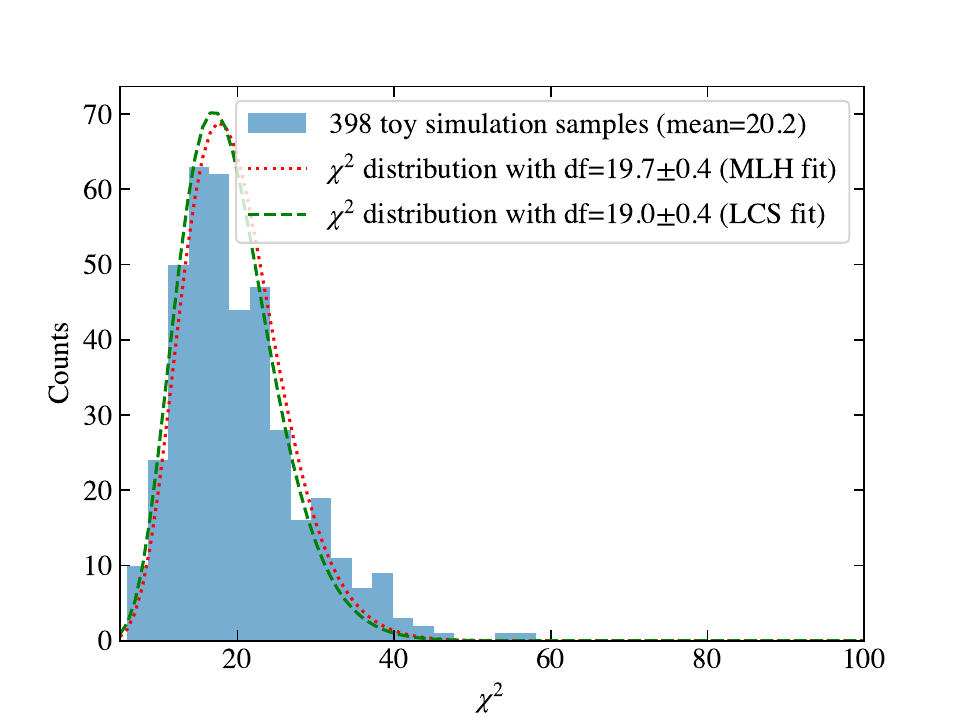}\put(-190,130){(b)}
    \caption{Toy studies using the measured cross sections of 400 toy fake data samples. The detailed descriptions of subplots (\textbf{a},\textbf{b}) are the same as described in Figure~\ref{fig:toy_validation_true}.}
    \label{fig:toy_validation_meas}
\end{figure}

In summary, a method as well as the corresponding procedures for the hadron-argon cross section measurement in an LArTPC detector were provided in this paper. The method requires the inputs of the initial energy and the energy at the end vertex of the track, as well as whether it is signal interaction occurring at the end vertex. The method showed good statistical performance, with no obvious bias except for that caused by unfolding, and good estimation of statistical uncertainties, as suggested by the toy studies. To apply it to real data, the systematic uncertainties due to the difference between the data and simulation should be considered, and the parameters of the unfolding algorithm used should be optimized with further investigations into the trade-off between bias and variance. These features can be added to the IPython notebook \texttt{hadron-Ar\_XS}~\cite{hadron-Ar_XS}, which also has the potential to be extended to more cross section studies.

\vspace{6pt} 




\funding{This material is based upon work supported by the National Science Foundation under Grant No. 2209601. }

\dataavailability{The notebook can be found at~\cite{hadron-Ar_XS}.} 



\acknowledgments{This study was initiated from analyzing pion-argon data on the ProtoDUNE-SP experiment~\cite{DUNE:2020cqd}. The author would like to thank Edward Blucher and Tingjun Yang for direct discussions on this study and also would like to thank Jacob Calcutt, Richard Diurba, Stephen Dolan, Elise Hinkle, Thomas Junk, Heng-Ye Liao, Laura Munteanu, Sungbin Oh, Ajib Paudel, Francesco Pietropaolo, Xin Qian, David Schmitz, Francesca Stocker, Leigh Whitehead, Kang Yang, and other collaborators for various useful discussions on the ProtoDUNE-SP data analyses.}

\conflictsofinterest{{The authors declare no conflict of interest. }}


%

\begin{adjustwidth}{-\extralength}{0cm}

 \reftitle{References}


\PublishersNote{}
\end{adjustwidth}
\end{document}